\documentclass[twocolumn,aip,nofootinbib,jcp,reprint,amsmath,amssymb]{revtex4-1}

\usepackage{natbib,hyperref}
\usepackage{amsmath}
\usepackage{graphicx}
\usepackage{hyperref} 
\usepackage{dcolumn}
\usepackage{bm}
\usepackage{epsfig,color,xspace,multirow,xr,bbold}
\usepackage[all]{xy}
\usepackage{setspace}
\usepackage{url}
\usepackage{soul}
\usepackage[colorinlistoftodos]{todonotes}
\usepackage{longtable}
\usepackage{tabularx}
\usepackage{adjustbox}
\usepackage{threeparttable} 
\usepackage{natbib,hyperref}
\usepackage{float}
\usepackage{placeins}
\usepackage[utf8]{inputenc}
\usepackage[T1]{fontenc}
\usepackage{array}

\usepackage[normalem]{ulem}
\usepackage{xcolor}

\newcommand\notsotiny{\@setfontsize\notsotiny\@vipt\@viipt}
\usepackage{soul} 
\newcommand{\orcid}[1]{\href{https://orcid.org/#1}{\includegraphics[width=8pt]{orcid.png}}}
\begin{document}

\title{
Intermolecular Packing and Cooperative Stabilization in Nucleotide Monophosphate Crystals
}

\date{\today}

\author{
Parimala Devi Duraisamy, 
Raghunathan Ramakrishnan$^*$
}

\email{ramakrishnan@tifrh.res.in}

\thanks{\\$^*$ Corresponding author}
\affiliation{$^1$Tata Institute of Fundamental Research, Hyderabad 500046, India}

\begin{abstract}
We examine the known polymorphs of cytidine 5$^\prime$-monophosphate (CMP) using dispersion-corrected density functional theory. 
The monoclinic phase is predicted to be the most stable, lying 1.76 and 3.00 kcal/mol per molecular unit below the orthorhombic and triclinic forms, respectively. 
Monoclinic CMP contains a ribose-based H-bonding motif along the crystallographic $\vec{b}$  direction, comprising alternating intra- and intermolecular interactions, motivating an investigation of cooperative stabilization along this pathway.
We employ chain-of-molecules (COM) and chain-of-slabs (COS) models to quantify the direction- and size-dependent stabilization and extend the analysis to the corresponding 5$^\prime$-monophosphates of thymidine, uridine, adenosine, and guanosine. 
Successive dimer-addition energies reveal the strongest size-dependent cooperative stabilization along the $\vec{b}$ direction, whereas growth along $\vec{a}$ and $\vec{c}$ exhibits substantially weaker size dependence. 
Among the nucleotides examined, adenosine 5$^\prime$-monophosphate shows the strongest chain-growth stabilization in the COM model, whereas the uridine variant shows the strongest stabilization in the COS model. Guanosine 5$^\prime$-monophosphate exhibits the weakest response in both models.
Minimal-basis iterative stockholder (MBIS) analysis further supports molecular polarization along the growing assemblies. 
These results 
demonstrate how nucleobase identity and the surrounding crystal environment modulate
direction-dependent cooperative stabilization associated with extended hydrogen-bond networks.
\end{abstract}




\maketitle




\section{Introduction}
\label{introduction}

Molecular crystals exhibiting polymorphism can adopt different packing
arrangements and form stable phases in distinct crystal structures. 
Intermolecular interactions, particularly hydrogen-bonding
(H-bonding) networks, can modulate the relative stability of 
various polymorphic structures.
\cite{vippagunta2001crystalline,cruz2014conformational,bernstein2020polymorphism}
Ab initio determination of the stability rankings of polymorphs is challenging because competing structures are often energetically separated by only a few
kJ/mol.\cite{hollingsworth2002crystal} Hence, an accurate
description of weak intermolecular interactions and their collective effects is essential for understanding the energetic origin of polymorphic
stability. Although empirical force fields have been successfully applied
to many organic molecular crystals, their transferability can be limited
when electronic effects dominate.
\cite{price2014predicting,della2024accurate} 
First-principles calculations,
particularly dispersion-corrected density functional theory (DFT), therefore
provide an important framework for examining the structures and relative
energies of molecular crystal polymorphs.
\cite{hunnisett2024seventh,nickerson2025assessment}

\begin{figure}[!hbpt]
    \centering
    \includegraphics[width=\columnwidth]{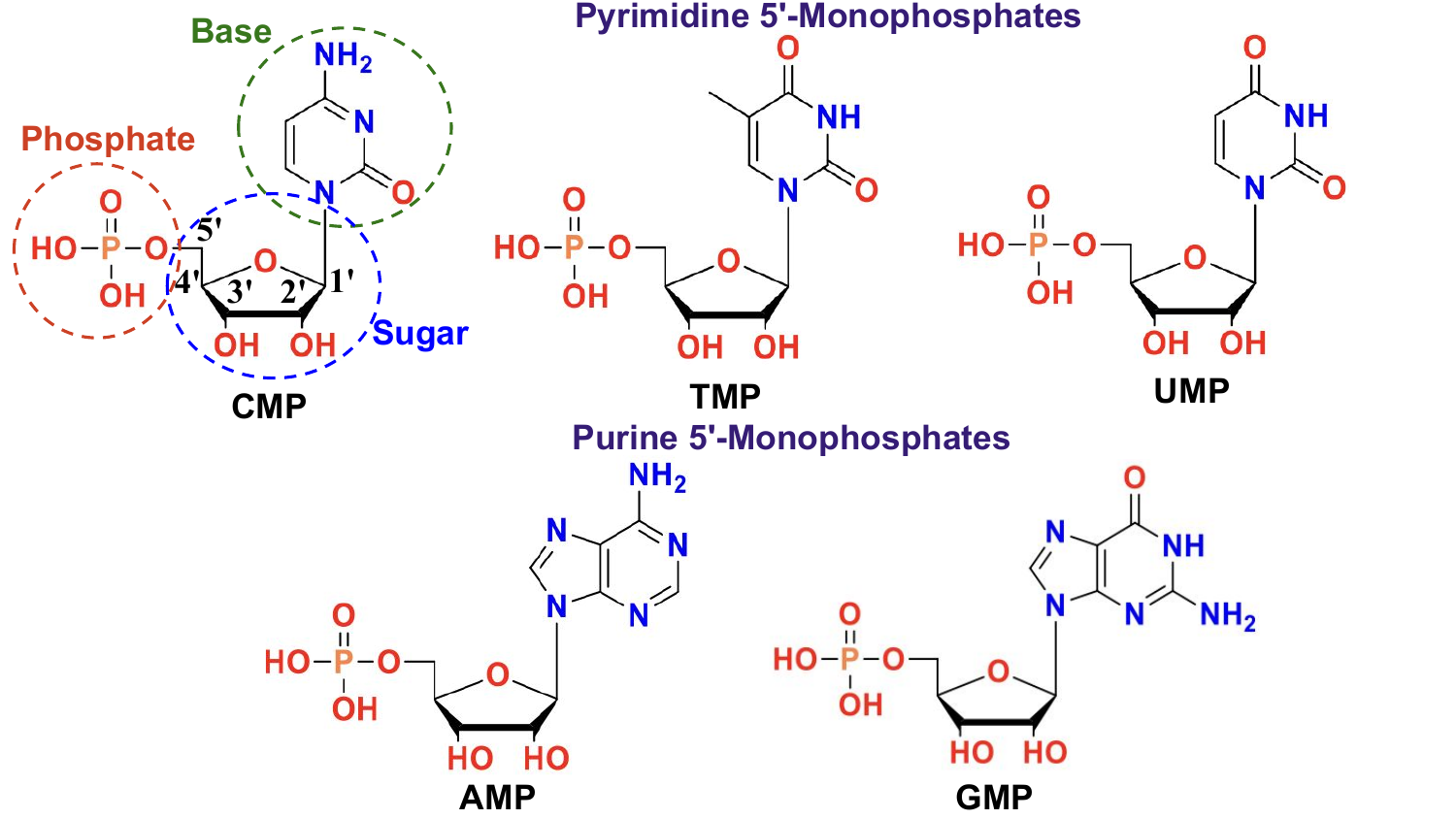}
    \caption{
    Nucleotide 5$^\prime$-monophosphates studied in this work.  
    }
    \label{fig:Molecular_Structure}
\end{figure}

Nucleotide 5$^\prime$-monophosphates (NMPs) contain a heterocyclic nucleobase, a ribose or deoxyribose sugar, and a phosphate group, providing multiple complementary hydrogen-bond donors and acceptors. Their distribution across the molecule enables extended intermolecular networks involving base, sugar, and phosphate groups, thereby supporting three-dimensional crystal packing.\cite{voet1970crystal, kartha1973cytidylic} 
The combination of nucleobase (pyrimidines: cytosine, thymine, and uracil; purines: adenine and guanine), sugar, and phosphate functionalities shown in Fig.~\ref{fig:Molecular_Structure} generates a diverse set of donor-acceptor patterns, enabling multiple competing H-bonding motifs, which makes the resulting crystal packing intrinsically nontrivial.

Cytidine 5$^\prime$-monophosphate (CMP) has been characterized experimentally in several solid-state forms, including anhydrous and hydrated phases.\cite{ sugawara2002crystallographic, borodi2001hydrated, gonzalez2005disodium}
More recently, three anhydrous CMP polymorphs with triclinic, monoclinic, and orthorhombic crystal structures were reported.\cite{kaszubowski2025same}
Their distinct intermolecular packing arrangements and H-bonding motifs offer an opportunity to examine the microscopic interactions that govern their relative stability.

A prominent structural feature of the monoclinic CMP phase is an extended
molecular packing motif propagating along the crystallographic $\vec{b}$
direction. Successive NMP units form a two-molecule zig-zag arrangement, with
the molecules occupying alternating positions on opposite sides of the chain.
Within this packing motif, the ribose groups participate in alternating intra-
and intermolecular O--H$\cdots$O interactions, while the nucleobase and
phosphate regions also form a recurring base--phosphate--base--phosphate
arrangement along the same direction.

Such an extended motif provides a structural basis for examining whether
stabilization changes nonadditively as the assembly grows and how this response
is modified by the surrounding crystal environment. Cooperative H-bonding has
been extensively studied in finite molecular clusters and aggregates,
\cite{dannenberg2002cooperativity,xantheas2000cooperativity,
tupikina2026cooperativity} but in crystals the energetic response
can also contain electrostatic, polarization, dispersion, and other
intermolecular contributions associated with the wider packing environment.
Hence, the NMP crystals provide a useful framework for examining
size-dependent stabilization of a propagating intermolecular motif and its
coupling to the lateral crystal environment.

Adenosine 5$^\prime$-monophosphate (AMP) is also experimentally known in a monoclinic crystal structure,
\cite{kaszubowski2025post} providing a purine-based example with an extended
packing motif along the corresponding chain-growth direction. Replacement of
the nucleobase while retaining a related ribose--phosphate framework provides
a means of examining how nucleobase chemistry modifies this molecular
arrangement. Differences in donor and acceptor functionality, nucleobase size,
and base--phosphate contacts can reorganize both the local H-bonding pattern
and the surrounding packing environment. CMP and AMP therefore provide
experimentally characterized reference structures, while TMP, UMP, and GMP
provide chemically distinct analogues for examining how the balance between
chain propagation and lateral crystal interactions changes across the NMP
series.

In this work, we first examine the experimentally reported anhydrous CMP
polymorphs using dispersion-corrected DFT, assess their relative stabilities,
and identify the extended molecular packing motif of monoclinic CMP along
$\vec{b}$. We then extend this structural framework to 
5$^\prime$-monophosphates of other nucleotides to examine how nucleobase identity modifies the
propagating intermolecular motif and its surrounding crystal environment.
To probe direction- and size-dependent stabilization, we employ complementary
chain-of-molecules (COM) and chain-of-slabs (COS) models along the three
crystallographic directions. Comparison of these models distinguishes the
stabilization of finite molecular assemblies from the additional influence of
the lateral crystal environment. Particular attention is given to the
$\vec{b}$ direction, where the NMPs form the extended zig-zag molecular
packing motif. The accompanying local electronic response is examined using
MBIS fragment charges and dipole-magnitude changes.

\begin{figure*}[!htpb]
\centering
\includegraphics[width=\textwidth]{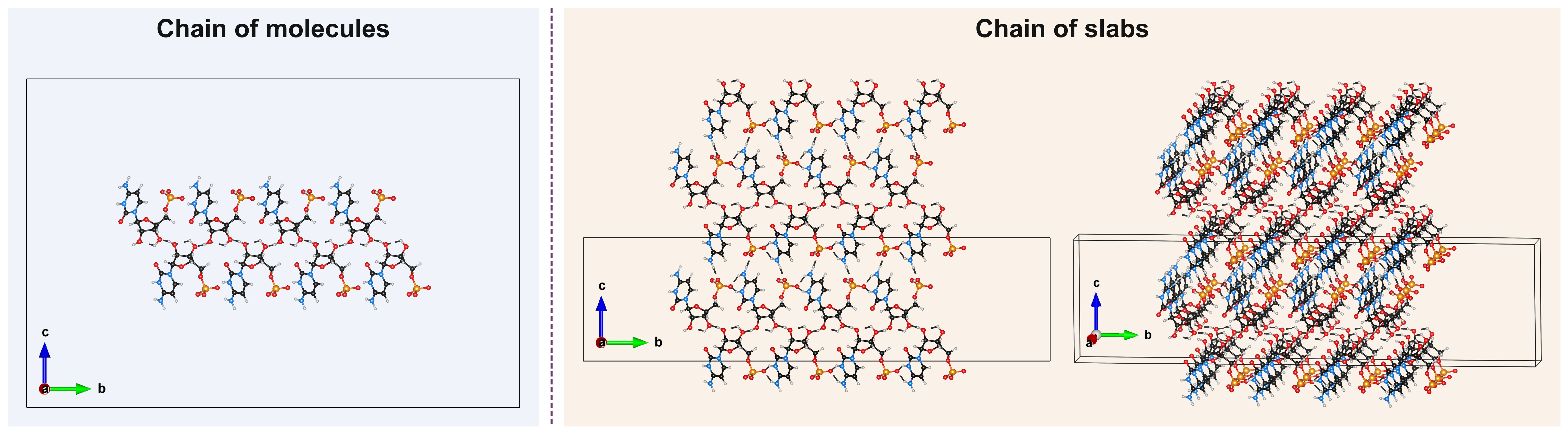}
\caption{
Structural models used to analyze the size-dependent stabilization of
nucleotide 5$^\prime$-monophosphate assemblies, illustrated using CMP.
Left: chain-of-molecules (COM) model, consisting of a finite segment of the
zig-zag molecular packing motif along $\vec{b}$. In CMP, this motif includes
alternating intra- and intermolecular ribose O--H$\cdots$O interactions
together with the recurring arrangement of nucleobase and phosphate regions.
Right: chain-of-slabs (COS) model, in which growth remains finite along
$\vec{b}$ while periodicity in the perpendicular $\vec{a}$ and $\vec{c}$
directions retains the lateral crystal environment.
}
\label{fig:NMP_COM_COS_models}
\end{figure*}

\section{Computational details}

Periodic density functional theory (DFT) calculations were performed 
using the Vienna
Ab initio Simulation Package (\textsc{VASP}).
\cite{kresse1996efficient,kresse1996efficiency}
The projector augmented-wave (PAW) method was used to describe the
interaction between the valence electrons and ionic cores,
\cite{blochl1994projector} together with a plane-wave kinetic-energy cutoff
of 500~eV. Brillouin-zone integrations
were performed using $\Gamma$-centered Monkhorst--Pack meshes\cite{monkhorst1976special} with a
reciprocal-space resolution of approximately 0.3~\AA$^{-1}$.
Structural optimizations were continued until the Hellmann--Feynman forces
on each atom were smaller than $5\times10^{-3}$~eV/\AA{} and the total
electronic energy was converged to within $1\times10^{-5}$~eV.

To assess the sensitivity of the CMP crystal structures to the treatment of exchange, correlation, and dispersion interactions within the DFT approximations, we used
experimental crystal structures of CMP polymorphs as benchmark data. 
Among the DFT approximations, we considered 
the generalized-gradient approximation (GGA)
functionals: PBE\cite{PhysRevLett.77.3865} and PBEsol,
\cite{perdew2008restoring} the dispersion-corrected PBE-D3(BJ)
method,\cite{PhysRevLett.77.3865,grimme2010consistent,grimme2011effect}; 
the meta-GGA methods: 
$r^{2}$SCAN+D3\cite{furness2020accurate,grimme2010consistent} and
SCAN+rVV10 methods,\cite{sun2015strongly,peng2016versatile}; and the
nonlocal van der Waals functionals: vdW-DF,\cite{dion2004van}
rev-vdW-DF2,\cite{hamada2014van} vdW-DF-CX,\cite{berland2014exchange}
and optB86b-vdW.\cite{klimevs2011van}
The selection of these methods was motivated by previous assessments of these
DFT approximations for molecular crystals and related
non-covalently bonded systems.
\cite{zhugayevych2023benchmark,della2025accurate,
berland2014van,peng2019van,bjorkman2014testing}
We compared the DFT-optimized lattice parameters of the CMP polymorphs with experimental values 
to identify a suitable DFT approximation for subsequent analysis across the NMP series, 
comprising CMP, thymidine 5$^\prime$-monophosphate (TMP), uridine 5$^\prime$-monophosphate (UMP), AMP, 
and guanosine 5$^\prime$-monophosphate (GMP).

We quantified the stability of the crystalline NMP structures using the
cohesive energy per NMP formula unit defined as
\begin{equation}
E_{\mathrm{coh}}
=
\frac{
E_{\mathrm{crystal}}
-
Z E_{\mathrm{vac}}(\mathrm{NMP})
}{Z},
\label{eq:cohesive_energy}
\end{equation}
where $E_{\mathrm{crystal}}$ is the total electronic energy of the periodic
crystal, $Z$ is the number of NMP formula units contained in the unit 
cell, and $E_{\mathrm{vac}}(\mathrm{NMP})$ is the energy of the
corresponding isolated NMP calculated in cubic cells with 40~\AA{} vacuum in all three directions. 
Accordingly, a negative value of
$E_{\mathrm{coh}}$ represents stabilization of the crystalline phase
relative to the isolated NMP unit.
Normalization by $Z$ in Eq.~\ref{eq:cohesive_energy} enables 
 comparison of CMP polymorphs having different numbers of formula units.

To examine the directional stabilization of the NMP crystal structures, we
constructed chain-of-molecules (COM) and chain-of-slabs (COS) models
independently along each of the three crystallographic directions,
$\vec{a}$, $\vec{b}$, and $\vec{c}$.
Fig.~\ref{fig:NMP_COM_COS_models} illustrates the models constructed along
$\vec{b}$. Both were generated directly from the corresponding optimized
crystal structures so that the molecular geometries and intermolecular
packing inherited from the parent crystal were retained.

Growth along $\vec{b}$ follows the extended zig-zag molecular arrangement
identified in the monoclinic CMP structure. In CMP this arrangement contains
the alternating ribose O--H$\cdots$O motif, but it also propagates the
relative arrangement of the nucleobase and phosphate regions. The
$\vec{b}$-direction models should therefore be regarded as finite segments of
the complete molecular packing motif rather than as isolated representations
of a single H-bond chain.

For each crystallographic direction, the dimer was taken as the fundamental
starting unit, and COM assemblies containing $N=2$, 4, 6, and 8 NMP units
were generated by successive addition of complete dimers along the selected
axis. This choice preserves the two-molecule repeat required across the
series: CMP, TMP, and AMP retain monoclinic $P2_1$ symmetry, whereas UMP
and GMP relax to triclinic $P1$ structures with two geometrically
inequivalent molecular environments arising from internal reorganization.
Using the dimer as the reference therefore avoids assuming equivalence of
individual monomers in the symmetry-lowered structures.
We introduced a 20~\AA{} vacuum separation in all three directions to suppress interactions between periodic images, so the COM systems represent isolated finite molecular chains. 
Along the growth direction, we adjusted the simulation-cell length for each chain size to maintain 
approximately 20~\AA{} of vacuum between neighboring periodic images.

We constructed the COS models analogously along each of the three crystallographic directions, retaining the periodic crystal environment in the two directions perpendicular to chain growth. 
Accordingly, a COS expanded along $\vec{a}$ is finite along $\vec{a}$ and periodic along $\vec{b}$ and $\vec{c}$, and so on.
We introduced a vacuum separation of 20~\AA{} only along the finite growth direction.

To examine how the local electronic environment responds to increasing chain
length, atomic charges and dipole contributions were obtained using the
minimal-basis iterative stockholder (MBIS) partitioning scheme with
ORCA 6.0.0~\cite{neese2012orca,neese2025software}, using PBE-D3(BJ),
the def2-TZVP basis set, and the RIJCOSX approximation with the def2/J
auxiliary basis.
\cite{weigend2005balanced,neese2009efficient,weigend2006accurate}

For the finite molecular chains, the response was monitored for a fixed
interior NMP or, where required by the symmetry-lowered packing, a central
two-NMP motif. Fragment charges and dipole magnitudes were used to determine
how strongly this local region continues to respond as increasingly distant
parts of the molecular assembly are introduced.

\section{Results and Discussion} 

\subsection{DFT benchmark for CMP polymorphs}

\begin{figure}[!hbpt]
    \centering
    \includegraphics[width=\columnwidth]{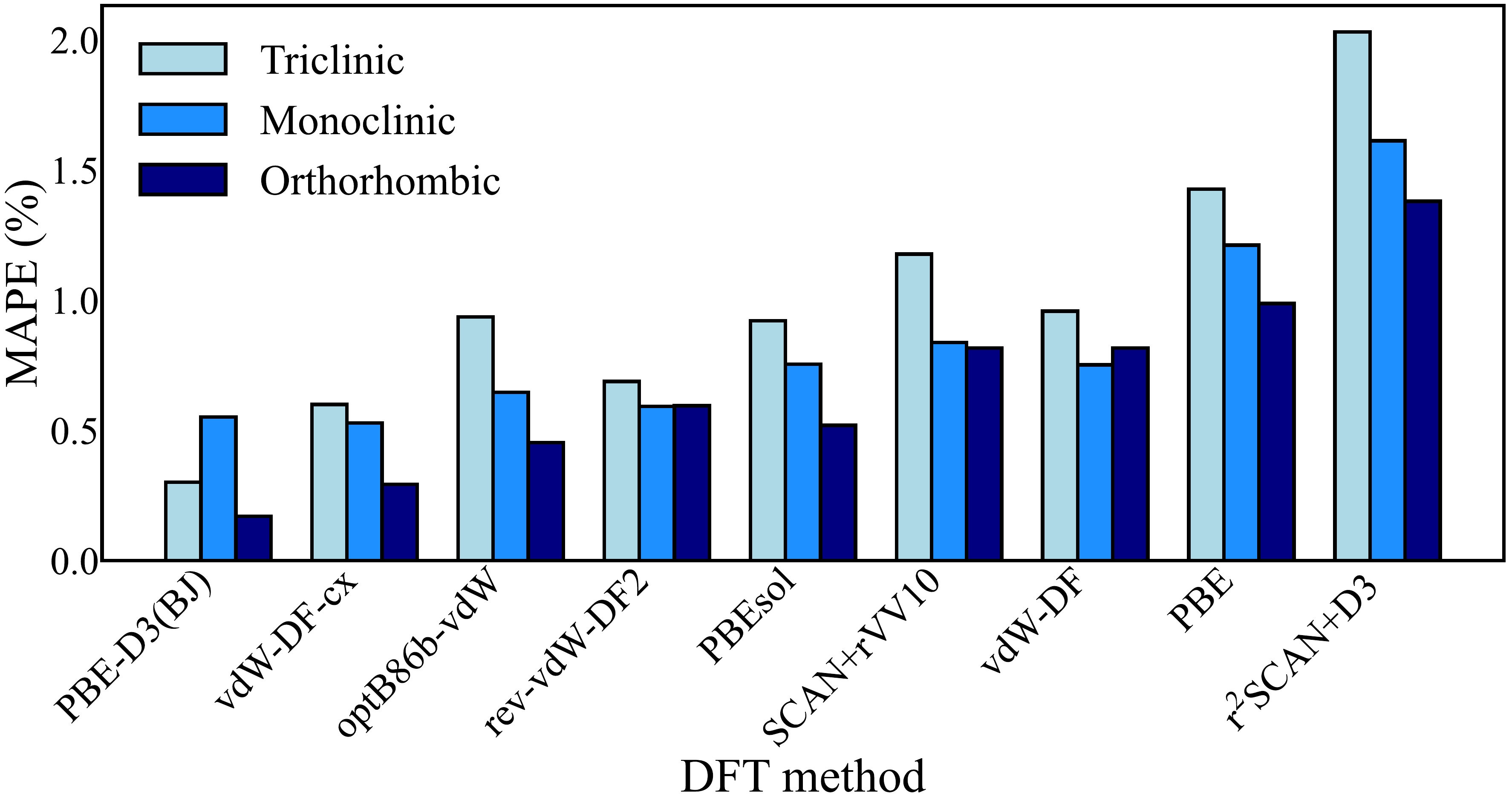}
    \caption{Mean absolute percentage error (MAPE) for   
    six lattice parameters ($a$, $b$, $c$, $\alpha$, $\beta$, and $\gamma$) of 
    triclinic, monoclinic, and orthorhombic CMP polymorphs predicted 
    using various DFT methods. The MAPE was determined by comparing the DFT-optimized lattice parameters with the corresponding experimental values.
    }
    \label{fig:CMP_DFT_benchmark}
\end{figure}

As molecular-crystal structures and relative polymorph stabilities are
sensitive to the treatment of weak intermolecular interactions\cite{otero2012benchmark}, particularly
dispersion, we benchmarked the DFT approximations considered here against the
experimentally reported triclinic, monoclinic, and orthorhombic CMP
polymorphs.
For each optimized structure, the lattice constants ($a$, $b$, and $c$) and
lattice angles ($\alpha$, $\beta$, and $\gamma$) were compared with experiment.
The overall performance was quantified using the mean absolute error (MAE),
mean absolute percentage error (MAPE), and standard deviation (SD).
The MAPE values are shown in Fig.~\ref{fig:CMP_DFT_benchmark}, the corresponding
MAE and SD values in Fig.~S1, and the individual lattice parameters in
Table~S1 of the supporting information (SI).

\begin{table*}[!hptb]
\centering
\caption{
Experimental crystal structural parameters for the
triclinic, monoclinic, and orthorhombic CMP polymorphs
along with those calculated with DFT-PBE-D3(BJ).
Relative energy ($\Delta E$) is reported per formula unit, with the
monoclinic polymorph taken as the reference.
Lattice constants ($a$, $b$, and $c$),
lattice angles ($\alpha$, $\beta$, and $\gamma$),
unit-cell volume ($V$), and $\Delta E$ are stated in
\AA, degrees, \AA$^3$, and
kcal/mol, respectively.
}
\label{tab:CMP_lattice}

\renewcommand{\arraystretch}{1.2}

\begin{tabular}{lll lll lll c}
\hline
\multicolumn{1}{l}{Structure} &
\multicolumn{1}{l}{Method~~~~~~~~~~} &
\multicolumn{1}{l}{$a$~~~~~~~~~~} &
\multicolumn{1}{l}{$b$~~~~~~~~~~} &
\multicolumn{1}{l}{$c$~~~~~~~~~~} &
\multicolumn{1}{l}{$\alpha$~~~~~~~~~~} &
\multicolumn{1}{l}{$\beta$~~~~~~~~~~} &
\multicolumn{1}{l}{$\gamma$~~~~~~~~~~} &
\multicolumn{1}{l}{$V$~~~~~~~~~~} &
\multicolumn{1}{l}{$\Delta E$} \\
\hline

Triclinic $(Z=1)$
& Exp.~\cite{kaszubowski2025same}
& 4.7896 & 8.1312 & 8.4477
& 76.85 & 79.75 & 83.99
& 314.55 & -- \\

& PBE-D3(BJ)
& 4.76 & 8.11 & 8.41
& 76.9 & 80.0 & 84.2
& 311.1 & 3.00 \\
\hline

Monoclinic $(Z=2)$
& Exp.~\cite{kaszubowski2025same}
& 4.843 & 8.646 & 15.726
& 90 & 93.68 & 90
& 657.1 & -- \\

& PBE-D3(BJ)
& 4.82 & 8.57 & 15.86
& 90.0 & 92.7 & 90.0
& 654.1 & 0.00 \\
\hline

Orthorhombic $(Z=4)$
& Exp.~\cite{kaszubowski2025same}
& 8.338 & 31.318 & 4.8912
& 90 & 90 & 90
& 1277.2 & -- \\

& PBE-D3(BJ)
& 8.33 & 31.36 & 4.85
& 90.0 & 90.0 & 90.0
& 1267.3 & 1.76 \\
\hline

\end{tabular}
\end{table*}

PBE-D3(BJ) provides the most consistent overall agreement with experiment
across the three CMP polymorphs. The predicted unit-cell volumes differ from
experiment by approximately 1.1\%, 0.5\%, and 0.8\% for the triclinic,
monoclinic, and orthorhombic forms, respectively, while the lattice constants
and angles show similarly small deviations (Table~\ref{tab:CMP_lattice}).
For the triclinic and orthorhombic polymorphs, PBE-D3(BJ) gives the lowest
MAE, MAPE, and SD; for monoclinic CMP, it gives the lowest MAE and SD,
whereas vdW-DF-CX gives the lowest MAPE. In general, the dispersion-corrected
methods outperform uncorrected PBE, with $r^2$SCAN+D3 as a notable exception
because of its tendency to produce more contracted lattice dimensions.~\cite{zhugayevych2023benchmark}
PBE-D3(BJ) was therefore selected for the subsequent structural, energetic,
and finite-assembly analyses.

The PBE-D3(BJ) results establish a clear energetic ordering among the three
CMP polymorphs. Using cohesive energies per formula unit, with the monoclinic
phase as the reference,
\begin{equation}
\Delta E(\mathrm{phase})=E_{\mathrm{coh}}(\mathrm{phase})-E_{\mathrm{coh}}(\mathrm{mono}),
\label{eq:relative_cohesive_energy}
\end{equation}
the orthorhombic and triclinic structures lie 1.76 and
3.00~kcal/mol per CMP above the monoclinic phase, respectively.

The molecular conformation is largely preserved across the three polymorphs.
The principal differences therefore arise
from crystal packing and from the topology of the intermolecular H-bonding
network.
The triclinic polymorph contains four intermolecular H-bonds per CMP,
involving ribose--phosphate and cytosine-mediated contacts. 
The orthorhombic polymorph contains five intermolecular H-bonds per CMP, including
ribose--base, ribose--ribose, and cytosine--phosphate interactions. 
Despite this larger number of H-bonds, the orthorhombic form is less stable than the
monoclinic structure, indicating that polymorph stability is not determined
by H-bond count alone.

\begin{figure*}[!hbpt]
\centering
\includegraphics[width=\textwidth]{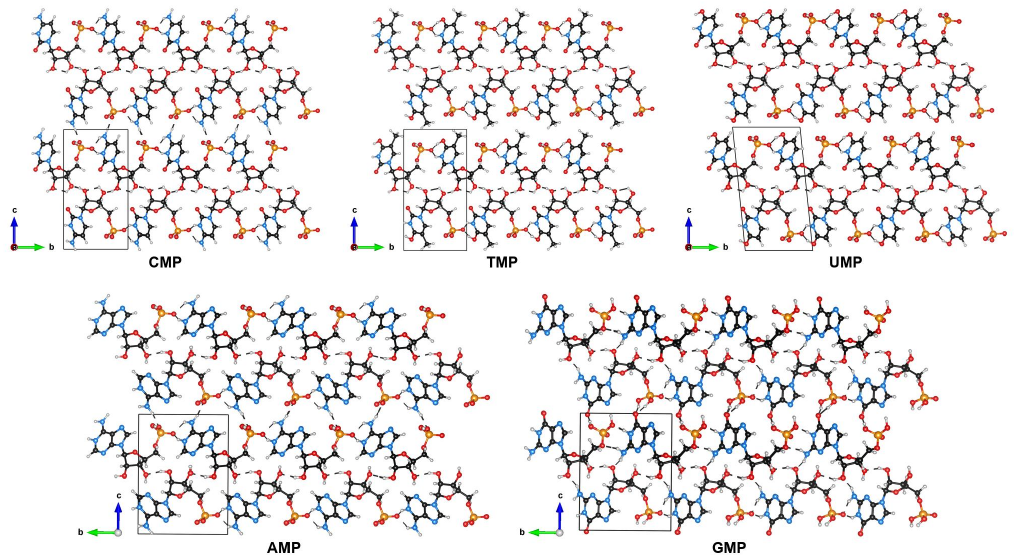}
\caption{
Supercell representations of the PBE-D3(BJ)-optimized anhydrous nucleotide
5$^\prime$-monophosphate crystal structures. 
}
\label{fig:CMP_AMP_structure}
\end{figure*}

The unit cell of monoclinic CMP contains four intermolecular H-bonds together with an
intramolecular O--H$\cdots$O interaction in the ribose region as shown in Fig.~\ref{fig:CMP_AMP_structure}. Along
$\vec{b}$, successive NMP units form an extended two-molecule zig-zag
arrangement. The ribose units participate in alternating intra- and
intermolecular O--H$\cdots$O interactions, while the nucleobase and phosphate
regions also alternate along the same propagation direction. 

This motif provides a natural structural coordinate for examining whether
stabilization evolves systematically with molecular assembly size and how the
surrounding lateral crystal environment modifies that behavior.

\begin{table*}[!hbpt]
\centering
\caption{
Experimental crystal structural parameters of NMP 
along with those calculated with DFT-PBE-D3(BJ).
CMP and AMP retain the experimentally observed $P2_1$ monoclinic structure. TMP, UMP, and GMP were generated from the corresponding CMP- or AMP-based structural motifs and subsequently optimized.
 Lattice constants ($a$, $b$, and $c$),
 lattice angles ($\alpha$, $\beta$, and $\gamma$),
 unit-cell volume ($V$), and cohesive energy
($E_{\mathrm{coh}}$) are stated in
 \AA, degrees, \AA$^3$, and
 kcal/mol (per NMP formula unit), respectively.
}
\label{tab:NMP_structures}

\renewcommand{\arraystretch}{1.2}

\begin{tabular}{lll lll lll  lc}
\hline
\multicolumn{1}{l}{NMP} & 
\multicolumn{1}{l}{Method~~~~~~~~~~} &
\multicolumn{1}{l}{$a$~~~~~~~~~~} & \multicolumn{1}{l}{$b$~~~~~~~~~~} & \multicolumn{1}{l}{$c$~~~~~~~~~~} &
\multicolumn{1}{l}{$\alpha$~~~~~~~~~~} & \multicolumn{1}{l}{$\beta$~~~~~~~~~~} & \multicolumn{1}{l}{$\gamma$~~~~~~~~~~} &
\multicolumn{1}{l}{$V$~~~~~~~~~~} & \multicolumn{1}{l}{$E_{\mathrm{coh}}$~~~~~~~~~~}  & \multicolumn{1}{l}{Space group} \\
\hline

CMP
& Exp.~\cite{kaszubowski2025same}
& 4.843 & 8.646 & 15.726
& 90 & 93.68 & 90
& 657.1 & & $P2_1$ \\
& PBE-D3(BJ)
& 4.82 & 8.57 & 15.86
& 90.0 & 92.7 & 90.0
& 654.1 & $-119.71$ & $P2_1$ \\
\hline

TMP
& PBE-D3(BJ)
& 4.76 & 8.65 & 16.65
& 90.0 & 95.1 & 90.0
& 682.4 & $-126.12$ & $P2_1$ \\
\hline

UMP
& PBE-D3(BJ)
& 4.78 & 8.46 & 15.50
& 95.9 & 87.9 & 90.0
& 622.9 & $-125.59$ & $P1$ \\
\hline

AMP
& Exp.~\cite{kaszubowski2025post}
& 4.8241 & 10.280 & 13.711
& 90 & 91.02 & 90
& 679.8 & & $P2_1$ \\
& PBE-D3(BJ)
& 4.81 & 10.25 & 13.66
& 90.0 & 91.3 & 90.0
& 673.2 & $-135.23$ & $P2_1$ \\
\hline

GMP
& PBE-D3(BJ)
& 5.17 & 10.32 & 13.33
& 89.6 & 99.5 & 92.7
& 700.8 & $-105.83$ & $P1$ \\
\hline
\end{tabular}
\end{table*}

\subsection{Extension of the monoclinic packing motif to other nucleotide monophosphates}

The extended packing motif observed along $\vec{b}$ in monoclinic CMP prompts
consideration of how nucleobase identity modifies the same chain-growth
coordinate. AMP provides an experimentally characterized purine-based
reference because its anhydrous crystal also adopts monoclinic $P2_1$
symmetry. We therefore extended the analysis to TMP, UMP, AMP, and GMP.
Fig.~\ref{fig:CMP_AMP_structure} shows the optimized structures of all five NMPs, and
Table~\ref{tab:NMP_structures} summarizes their structural parameters,
cohesive energies, and space groups. All optimized coordinates are provided in the SI.

For the two systems with experimentally characterized monoclinic structures, CMP and AMP, PBE-D3(BJ) preserves the experimental $P2_1$ symmetry and closely reproduces the lattice parameters. 
Upon nucleobase substitution, TMP also retains the CMP-derived $P2_1$ packing, whereas UMP and GMP relax to symmetry-lowered $P1$ structures containing two inequivalent molecular environments. 

Nucleobase substitution also changes protonation and tautomeric state. 
TMP and UMP reorganize from the CMP-derived starting structures during optimization, while GMP undergoes proton transfer from guanine to the phosphate group. 
These changes modify the local base--phosphate interaction network and are particularly pronounced for GMP, which also shows the largest departure from the packing of the experimentally characterized CMP and AMP structures.

\begin{figure*}[!hbpt]
\centering
\includegraphics[width=\textwidth]
{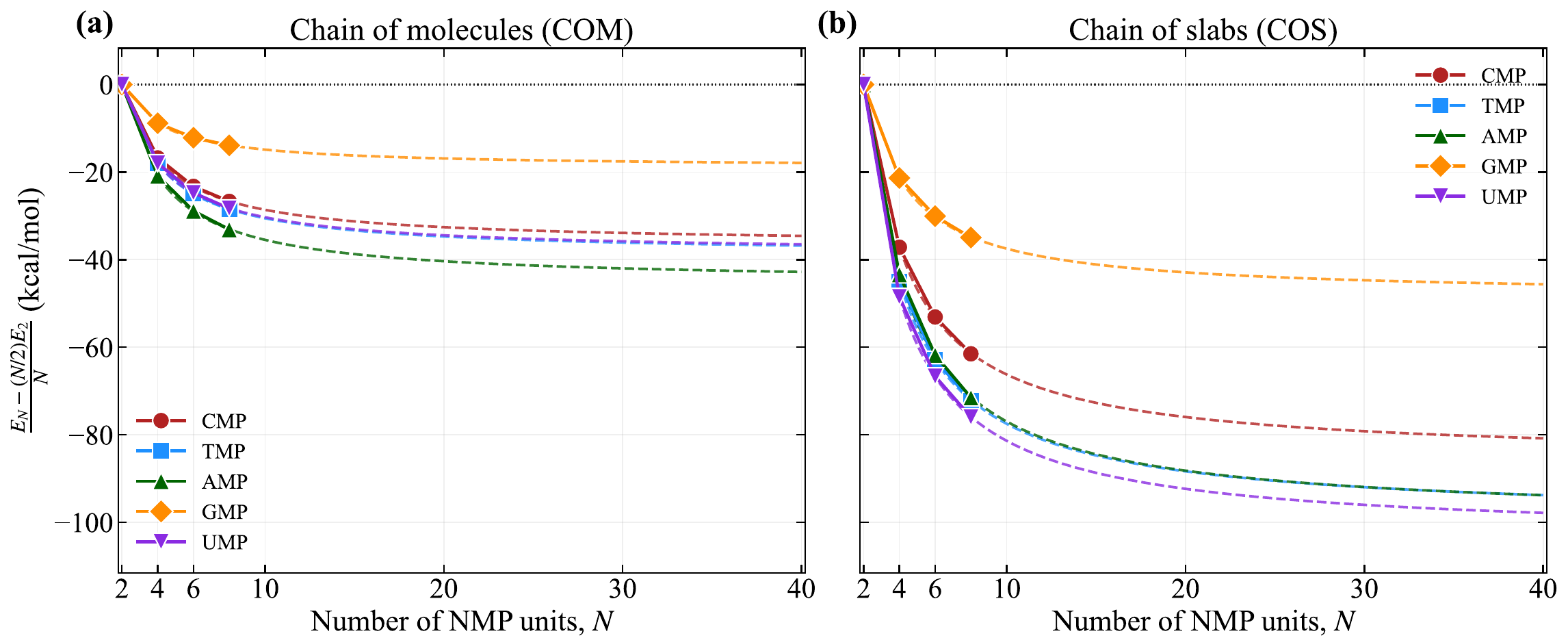}
\caption{
Stabilization energy per NMP formula unit as a function of chain length, $N$, for (a) chain-of-molecules (COM) and (b) chain-of-slabs (COS) models. 
Stabilization energy is defined as
$[E_N-(N/2)E_2]/N$, where $E_N$ is the energy of the $N$-unit assembly
and $E_2$ is the corresponding dimer reference energy. Symbols denote
the explicitly calculated values for $N=2,4,6,$ and $8$, while
dashed lines represent extrapolated fitted curves to
$\Delta_N=\Delta_\infty+A/N$.
}
\label{fig:dimer_reference_stabilization}
\end{figure*}

Despite differences in symmetry and local intermolecular geometry, all five
systems retain a propagating molecular arrangement along the selected
chain-growth direction. Yet, the detailed H-bonding pattern superimposed on this
packing motif depends on nucleobase identity. In CMP, UMP, and GMP,
the ribose units participate in an alternating sequence of intra- and
intermolecular O--H$\cdots$O interactions. In TMP and AMP, the corresponding
intramolecular ribose O--H$\cdots$O contact is absent; instead, the ribose
hydroxyl forms an intermolecular O--H$\cdots$N interaction with the
nucleobase of the neighboring NMP unit. The relevant intra- and intermolecular ribose hydrogen-bond geometries are summarized in Table ~S2. At the same time, interactions
involving the nucleobase and phosphate regions form part of the same extended
packing environment. Thus, the common coordinate across the NMP series is
the propagation of the molecular packing motif rather than an identical
sequence of H-bonds.

The cohesive energies in Table~\ref{tab:NMP_structures} vary substantially across the series, with AMP and GMP showing the largest and the smallest crystal stabilization, respectively. 
Since these values are referenced to the corresponding isolated NMPs, they reflect overall crystal stabilization rather than the strength of any individual H-bonding motif.

\begin{table*}[!hbpt]
\centering
\caption{Incremental stabilization energies associated with successive
addition of two-unit NMP motifs in the chain-of-molecules (COM) and
chain-of-slabs (COS) models. The first increment, $E_4-2E_2$, describes
formation of the four-unit assembly from two isolated dimers;
$E_6-E_4-E_2$ describes addition of a third dimer to the four-unit
assembly; and $E_8-E_6-E_2$ describes addition of a fourth dimer to the
six-unit assembly. Values in parentheses are the changes in the incremental stabilization energy between two consecutive additions, $\Delta\Delta E^{(2)}_{4/6}$ and $\Delta\Delta E^{(2)}_{6/8}$. All values are in kcal/mol. }
\label{tab:dimer_incremental_stabilization}
\renewcommand{\arraystretch}{1.15}
\begin{tabular}{lc rrr c rrr}
\hline
NMP & Direction
& \multicolumn{4}{l}{COM}
& \multicolumn{3}{l}{COS} \\
\cline{3-5}\cline{7-9}
&
& $E_4-2E_2~$
& $E_6-E_4-E_2$
& $E_8-E_6-E_2$
&~~~~~~
& $E_4-2E_2$
& $E_6-E_4-E_2$
& $E_8-E_6-E_2$ \\
\hline
CMP & $a$ &$-$30.83 &$-$30.58 ($+$0.25) & $-$30.13 ($+$0.45) &&$-$101.49 &$-$104.86 ($-$3.37)&$-$105.14 ($-$0.28)\\
    & $b$ &$-$67.02 &$-$72.54 ($-$5.52) & $-$73.94 ($-$1.40)&& $-$148.65 &$-$169.95 ($-$21.30) &$-$173.52 ($-$3.57)\\
    & $c$ &$-$23.90 &$-$23.89 ($+$0.01)&$-$23.89 (~~0.00)&& $-$32.53 &$-$32.55 ($-$0.02)&$-$32.55 (~~0.00)\\
TMP & $a$ &$-$37.32 &$-$38.08 ($-$0.76)&$-$37.74 ($+$0.34)&&$-$151.62 & $-$153.28  ($-$1.66)& $-$153.61 ($-$0.33)\\
    & $b$ &$-$72.26 &$-$77.26 ($-$5.00)&$-$78.55 ($-$1.29)&& $-$179.83 &$-$197.29 ($-$17.46)&$-$200.31 ($-$3.02)\\
    & $c$ &$-$1.27 & $-$1.22 ($+$0.05)& $-$1.23 ($-$0.01)&&$-$8.10 &$-$8.17 ($-$0.07)& $-$8.18 ($-$0.01)\\
UMP & $a$ & $-$34.98 & $-$35.47 ($-$0.49)&$-$35.12 ($+$0.35)&&$-$23.35 &$-$24.87 ($-$1.52)&$-$24.97 ($-$0.10)\\
    & $b$ &$-$71.86 &$-$76.63 ($-$4.77)&$-$77.87 ($-$1.24)&&$-$194.05 &$-$205.89 ($-$11.84)& $-$207.90 ($-$2.01)\\
    & $c$ & 3.29 & 3.38 ($+$0.09)&3.36 ($-$0.02)&&14.33 &14.23 ($-$0.10)&14.12 ($-$0.11)\\
AMP & $a$ & $-$32.75 & $-$32.55 ($+$0.20)& $-$32.13 ($+$0.42)&&$-$158.03 &$-$165.21 ($-$7.18)& $-$165.64 ($-$0.43)\\
    & $b$ & $-$83.41 &$-$89.87 ($-$6.46)& $-$91.50 ($-$1.63)&&$-$173.52 &$-$197.34 ($-$23.82) &$-$201.23 ($-$3.89)\\
    & $c$ &5.94 &6.09 ($+$0.15)&6.08 ($-$0.01)&&$-$29.24  &$-$29.24  (~~0.00)& $-$29.34 ($-$0.10)\\
GMP & $a$ &$-$15.08 &$-$16.16 ($-$1.08)&$-$16.32 ($-$0.16)&&$-$37.57 &$-$42.66 ($-$5.09)&$-$43.23 ($-$0.57)\\
    & $b$ &$-$35.33 &$-$37.57 ($-$2.24)&$-$38.16 ($-$0.59)&&$-$85.36 &$-$95.01 ($-$9.65)&$-$99.18 ($-$4.17)\\
    & $c$ & $-$25.71 &$-$25.98 ($-$0.27)& $-$26.02 ($-$0.04)&& $-$96.08&$-$96.36 ($-$0.28)&$-$96.53 ($-$0.17)\\
\hline
\end{tabular}
\end{table*}

The NMP series therefore provides a common chain-growth coordinate while
retaining chemically distinct H-bonding, base--phosphate, and lateral
intermolecular environments. This allows the size-dependent stabilization of
the propagating molecular assemblies to be compared and the influence of the
surrounding crystal environment to be assessed.

\subsection{Size-dependent stabilization and cooperative growth of the NMP assemblies}

To probe whether stabilization becomes increasingly collective as the assemblies
grow beyond the dimer reference, we consider the energetic effect of successive
dimer addition. For an assembly containing an even number $N$ of NMP units,
\begin{equation}
\Delta E_{N}^{(2)}
=
E_{N}-\left(E_{N-2}+E_{2}\right),
\qquad N=4,6,8,\ldots
\label{eq:dimer_add_general}
\end{equation}
where $E_N$ is the total energy of the corresponding COM or COS assembly.
More negative values indicate increasingly favorable dimer addition.
The calculated values are summarized in
Table~\ref{tab:dimer_incremental_stabilization}, with the corresponding
total energies given in Table~S3.

Comparison of $\Delta E_N^{(2)}$ for $N=4$, 6, and 8 along the three
crystallographic directions shows the strongest size dependence along
$\vec{b}$. For all five NMPs, successive dimer additions along $\vec{b}$
become progressively more favorable, whereas the corresponding changes along
$\vec{a}$ and $\vec{c}$ are generally much smaller. The directional contrast
therefore identifies the propagating $\vec{b}$-direction packing motif as the
dominant coordinate for size-dependent stabilization.

For the isolated COM assemblies along $\vec{b}$, AMP exhibits the strongest
incremental stabilization, whereas GMP shows the weakest response. CMP, TMP,
and UMP display comparable intermediate behavior. 
In all cases, successive
extension remains energetically favorable, showing that the stabilization
associated with propagation of the molecular motif is not exhausted within
the initial dimer.
The COS assemblies show the same qualitative size dependence but substantially
more favorable dimer-addition energies than the corresponding COM chains.
The increasing COM--COS separation with assembly size shows that the lateral
crystal environment modifies the stabilization associated with propagation
along $\vec{b}$ rather than merely adding a constant energetic offset.

To compare the overall stabilization of assemblies of different size on a
common molecular basis, we define
\begin{equation}
\varepsilon_N^{\mathrm{NMP}}
=
\frac{E_N-(N/2)E_2}{N},
\label{eq:dimer_reference_per_nmp}
\end{equation}
which measures the additional stabilization per NMP relative to an equivalent
set of isolated dimers. The calculated values become progressively more
negative with increasing $N$ for both COM and COS models
(Fig.~\ref{fig:dimer_reference_stabilization}), consistent with the
incremental growth analysis.

\begin{table}[hpbt]
\centering
\caption{
Extrapolated dimer-referenced stabilization energies of NMPs according to
the chain-of-molecules (COM) and chain-of-slabs (COS) models.
The stabilization energy is extrapolated according to
$\Delta_N=\Delta_{\infty}+A/N$.
The lateral stabilization energy,
$\Delta_{\mathrm{lat}}
=
\Delta_{\infty}^{\mathrm{COS}}
-
\Delta_{\infty}^{\mathrm{COM}}$,
quantifies the additional stabilization gained when the one-dimensional
molecular chain is embedded in the surrounding crystal environment.
The ratio
$|\Delta_{\mathrm{lat}}/\Delta_{\infty}^{\mathrm{COM}}|$
gives the magnitude of this additional stabilization relative to the
COM stabilization.
Energies are given in kcal/mol per formula unit of NMP.
}
\label{tab:COM_COS_extrapolation}

\renewcommand{\arraystretch}{1.2}

\begin{tabular}{l rrr c}
\hline
\multicolumn{1}{l}{NMP~~~~~} &
\multicolumn{1}{l}{$\Delta_{\infty}^{\mathrm{COM}}$~~~~~} &
\multicolumn{1}{l}{$\Delta_{\infty}^{\mathrm{COS}}$~~~~~} &
\multicolumn{1}{l}{$\Delta_{\mathrm{lat}}$~~~~~} &
\multicolumn{1}{l}{$\left|\Delta_{\mathrm{lat}}/
\Delta_{\infty}^{\mathrm{COM}}\right|$~~~~} \\
\hline
CMP & $-36.55$ & $-85.68$  & $-49.13$ & 1.34 \\
TMP & $-38.88$ & $-99.24$  & $-60.36$ & 1.55 \\
UMP & $-38.56$ & $-103.34$ & $-64.78$ & 1.68 \\
AMP & $-45.25$ & $-99.43$  & $-54.18$ & 1.20 \\
GMP & $-18.90$ & $-48.33$  & $-29.42$ & 1.56 \\
\hline
\end{tabular}
\end{table}

The finite-size dependence was described using
\begin{equation}
\varepsilon_N^{\mathrm{NMP}}
=
\Delta_{\infty}
+
\frac{A}{N},
\label{eq:finite_size_extrapolation}
\end{equation}
where $\Delta_{\infty}$ is the extrapolated dimer-referenced stabilization per
NMP in the large-$N$ limit. The resulting values are summarized in
Table~\ref{tab:COM_COS_extrapolation}.

For the COM models, the magnitude of
$\Delta_{\infty}^{\mathrm{COM}}$ follows approximately
\begin{equation}
\mathrm{AMP}
>
\mathrm{TMP}
\approx
\mathrm{UMP}
\approx
\mathrm{CMP}
\gg
\mathrm{GMP}.
\end{equation}
AMP therefore shows the strongest stabilization associated with propagation
of the isolated molecular chain, while GMP remains substantially weaker.
CMP, TMP, and UMP form an intermediate group with similar large-$N$
stabilization.

The COS extrapolations are considerably more negative for every NMP.
UMP exhibits the strongest large-$N$ stabilization in the slab environment,
with TMP and AMP close behind, whereas GMP again shows the weakest response.
The change in ordering between COM and COS indicates that the relative
stabilization of the NMPs depends strongly on the surrounding lateral crystal
environment.

To quantify this effect, we define
\begin{equation}
\Delta_{\mathrm{lat}}
=
\Delta_{\infty}^{\mathrm{COS}}
-
\Delta_{\infty}^{\mathrm{COM}}.
\label{eq:lateral_cooperative_enhancement}
\end{equation}
The lateral contribution is substantial for all five NMPs and ranges from 1.20 to 1.68 times the magnitude of the corresponding COM stabilization. UMP shows the largest absolute lateral contribution, whereas GMP shows the smallest. 
Since the COS--COM difference contains direct lateral intermolecular
interactions together with changes in electrostatic, polarization, dispersion,
and other nonadditive contributions within the propagating assembly,
$\Delta_{\mathrm{lat}}$ should be interpreted as the net influence of the
surrounding crystal environment rather than as a separable lateral
interaction energy.

\subsection{Local electronic response along the propagating NMP chain}
\label{sec:dipole}

The energetic analysis shows that extension of the NMP assemblies along
$\vec{b}$ produces increasingly favorable stabilization. To examine how this
growth modifies the electronic structure of an interior region of the chain,
we analyzed the MBIS fragment charges and local dipole-magnitude response as
the surrounding NMP environment was progressively extended.

For a fragment $\mathcal{F}$, the net MBIS charge was obtained by summing the
atomic MBIS charges over all atoms assigned to that fragment,
\begin{equation}
q_{\mathcal{F}} = \sum_{A\in\mathcal{F}} q_A,
\label{eq:mbis_fragment_charge}
\end{equation}
where $q_A$ is the MBIS charge of atom $A$. The change associated with chain
extension from $N_1$ to $N_2$ was then defined as
\begin{equation}
\Delta q_{\mathcal{F}}(N_2-N_1)
= q_{\mathcal{F}}(N_2) - q_{\mathcal{F}}(N_1).
\label{eq:mbis_fragment_charge_change}
\end{equation}

The NMP was partitioned into nucleobase, phosphate, and sugar fragments.
The net MBIS charge of each fragment was obtained from
Eq.~\ref{eq:mbis_fragment_charge}, and the changes associated with successive
chain extension were evaluated using
Eq.~\ref{eq:mbis_fragment_charge_change}. The resulting values are summarized
in Table~\ref{tab:fragment_charge_changes}.

For CMP, TMP, and AMP, the dominant charge redistribution occurs upon
extension from $N=3$ to $N=5$. This step produces a pronounced compensating
response between the nucleobase and phosphate fragments, whereas the
subsequent $7-5$ and $9-7$ changes are much smaller. UMP shows an analogous
behavior, with a large $6-2$ response followed by only a small additional
change from $N=6$ to $N=10$. GMP exhibits substantially weaker fragment-charge
redistribution throughout. The near cancellation of the fragment-charge
changes in the total charge indicates that the response primarily involves
redistribution within the monitored molecular region rather than substantial
net charge transfer to or from it.

\begin{table}[!htpb]
\centering
\caption{
Changes in MBIS fragment charges upon extension of the NMP chains.
$\Delta q_{\mathrm{base}}$, $\Delta q_{\mathrm{phos.}}$, and
$\Delta q_{\mathrm{sugar}}$ denote changes in the net MBIS charges of the
base, phosphate, and sugar fragments for the given chain sizes.
$\Delta q_{\mathrm{total}}$ is the corresponding change in the charge of
the monitored central NMP or central $AB$ unit. Charges are given in units
of $e$.
}
\label{tab:fragment_charge_changes}

\renewcommand{\arraystretch}{1.15}

\begin{tabular}{llrrrr}
\hline
NMP&Extension & \multicolumn{4}{l}{$\Delta q$} \\
\cline{3-6}
 &
 &
\multicolumn{1}{l}{base} &
\multicolumn{1}{l}{phos.} &
\multicolumn{1}{l}{sugar} &
\multicolumn{1}{l}{total} \\
\hline

CMP
& $3-1$ & $-0.0288$ & $+0.0168$ & $+0.0091$ & $-0.0029$ \\
& $5-3$ & $+0.1144$ & $-0.0977$ & $-0.0120$ & $+0.0047$ \\
& $7-5$ & $+0.0011$ & $-0.0001$ & $-0.0018$ & $-0.0008$ \\
& $9-7$ & $+0.0011$ & $-0.0007$ & $-0.0008$ & $-0.0004$ \\
\hline

TMP
& $3-1$ & $-0.0287$ & $+0.0181$ & $+0.0073$ & $-0.0033$ \\
& $5-3$ & $+0.1385$ & $-0.1392$ & $+0.0047$ & $+0.0040$ \\
& $7-5$ & $+0.0014$ & $-0.0002$ & $-0.0018$ & $-0.0006$ \\
& $9-7$ & $+0.0006$ & $-0.0008$ & $+0.0002$ & $0.0000$ \\
\hline

AMP
& $3-1$ & $-0.0288$ & $+0.0406$ & $-0.0294$ & $-0.0177$ \\
& $5-3$ & $+0.1666$ & $-0.1454$ & $-0.0027$ & $+0.0185$ \\
& $7-5$ & $0.0000$  & $+0.0007$ & $-0.0017$ & $-0.0009$ \\
& $9-7$ & $+0.0006$ & $0.0000$  & $-0.0004$ & $+0.0002$ \\
\hline

UMP
& $6-2$  & $+0.2708$ & $-0.2757$ & $+0.0056$ & $+0.00035$ \\
& $10-6$ & $+0.0037$ & $-0.0020$ & $-0.0025$ & $-0.00040$ \\
\hline

GMP
& $6-2$  & $+0.0050$ & $-0.0003$ & $-0.0031$ & $+0.00085$ \\
& $10-6$ & $+0.0010$ & $-0.0036$ & $+0.0012$ & $-0.00070$ \\
\hline

\end{tabular}
\end{table}

The fragment-resolved response also shows that the electronic consequences
of chain growth are not confined to the ribose H-bonding region. For CMP,
TMP, AMP, and particularly UMP, the dominant redistribution occurs between
the nucleobase and phosphate fragments, while the sugar contribution is
generally smaller. This behavior is consistent with the broader
$\vec{b}$-direction packing motif, in which ribose-centered H-bonding and the
relative arrangement of nucleobase and phosphate groups propagate together.

The MBIS-derived dipole response provides a complementary measure of the
local polarization response of the same interior region.
For CMP, TMP, and AMP, odd-numbered chains with
$N=3$, 5, 7, 9, and 11 contain a unique central NMP whose position within the
chain remains unchanged as the surrounding environment is enlarged. For UMP
and GMP, the optimized $P1$ structures contain two inequivalent molecular
environments forming an $\cdots ABAB\cdots$ sequence. The central $AB$ motif
was therefore used as the local probe, with $N=2$, 6, and 10 selected so that
the same $AB$ ordering is retained.

\begin{figure*}[!hptb]
\centering
\includegraphics[width=\textwidth]{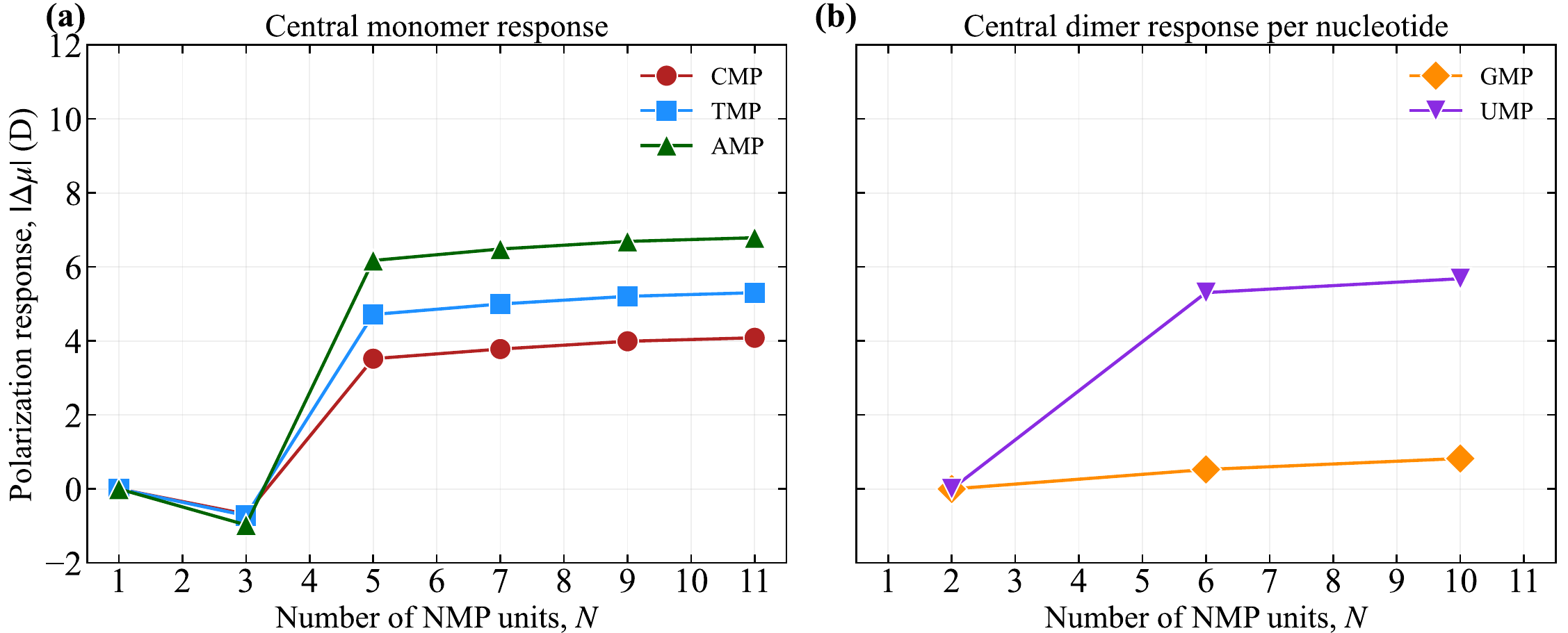}
\caption{
Chain-length dependence of the MBIS-derived local dipole-magnitude response
along the crystallographic $\vec{b}$ direction.
(a) For CMP, TMP, and AMP, $\Delta\mu_{\mathrm{mono}}$ is the change in the
central NMP dipole magnitude relative to the corresponding isolated monomer.
(b) For UMP and GMP, which contain alternating inequivalent molecular
environments, $\Delta\mu_{AB}$ is evaluated for the central $AB$ motif
relative to the isolated $AB$ dimer and reported per NMP.
}
\label{fig:NMPdipole}
\end{figure*}

The dipole of a selected fragment $\mathcal{F}$ was reconstructed from the
MBIS atomic charges and atom-centered dipoles according to
\cite{verstraelen2016minimal,mikkelsen2025minimal}

\begin{equation}
\boldsymbol{\mu}_{\mathcal{F}}
=
\sum_{A\in\mathcal{F}}
q_A
\left(
\mathbf{R}_A-\mathbf{R}_0
\right)
+
\sum_{A\in\mathcal{F}}
\boldsymbol{\mu}^{\mathrm{MBIS}}_A,
\label{eq:mbis_dipole}
\end{equation}

where $q_A$, $\mathbf{R}_A$, and
$\boldsymbol{\mu}^{\mathrm{MBIS}}_A$ are the MBIS charge, atomic position,
and atom-centered dipole of atom $A$, respectively. A common center-of-mass
origin was used for each monitored fragment. Applying the same origin
convention throughout is important because the MBIS partitioning can assign
a small residual charge to a molecular fragment within the interacting chain.

For CMP, TMP, and AMP, the local response was defined as the change in the
magnitude of the central-NMP dipole relative to the corresponding isolated
monomer,

\begin{equation}
\Delta\mu_{\mathrm{mono}}(N)
=
\left|
\boldsymbol{\mu}_{\mathrm{central}}(N)
\right|
-
\left|
\boldsymbol{\mu}_{1}
\right|.
\label{eq:delta_mu_monomer}
\end{equation}

For UMP and GMP, the corresponding quantity was evaluated for the central
$AB$ motif relative to the isolated $AB$ dimer and normalized per NMP,

\begin{equation}
\Delta\mu_{AB}(N)
=
\frac{
\left|
\boldsymbol{\mu}_{AB}^{\mathrm{central}}(N)
\right|
-
\left|
\boldsymbol{\mu}_{AB}(2)
\right|
}{2}.
\label{eq:delta_mu_dimer}
\end{equation}

These quantities describe changes in dipole magnitudes and should therefore
be distinguished from the magnitude of the vector difference
$|\boldsymbol{\mu}(N)-\boldsymbol{\mu}_{\mathrm{ref}}|$.

As shown in Fig.~\ref{fig:NMPdipole}, CMP, TMP, and AMP exhibit a pronounced
increase in the central dipole-magnitude response on extending the chain from
$N=3$ to $N=5$. The corresponding values increase from 0.68 to 3.52~D for
CMP, from 0.72 to 4.72~D for TMP, and from 0.97 to 6.18~D for AMP.
Subsequent extension produces much smaller changes, with the responses
reaching 4.09, 5.30, and 6.79~D, respectively, at $N=11$.

The $3\rightarrow5$ step is particularly informative because the central NMP
already has its directly interacting neighbors in the trimer. Extension to
the pentamer therefore introduces the next molecular shell without adding a
new direct H-bond to the central NMP. The pronounced change in its dipole
magnitude shows that the electronic influence of the surrounding chain extends
beyond the directly interacting nearest neighbors. 
The much smaller changes at larger $N$ indicate that the dipole response of
this fixed interior region approaches saturation once the second-neighbor
environment is established.

UMP and GMP show the same qualitative separation between the initial
environmental response and subsequent convergence. Embedding the central
$AB$ motif within a six-membered chain produces a response of 5.31~D per NMP
for UMP but only 0.53~D per NMP for GMP. Extension to $N=10$ increases these
values only to 5.68 and 0.82~D per NMP, respectively.

The rapid convergence of the fragment charges and dipole response describes
the behavior of a fixed interior region and should not be interpreted as
saturation of the stabilization of the growing assembly. Once the nearby
environment has been established, additional NMP units can interact with an
approximately converged local charge distribution while simultaneously
forming new intermolecular contacts elsewhere in the assembly. Energetic
stabilization can therefore continue even when the electronic state of the
central probe changes only weakly.

This distinction is particularly important for the finite COM and COS models.
Both remain terminated along the growth direction. A finite COM assembly does
not recover the large-$N$ stabilization of the corresponding infinite isolated
molecular chain, whereas a finite COS assembly retains the lateral periodic
environment but does not recover the limit in which propagation is also
restored along the growth direction. The continued evolution of the
finite-system energies and their large-$N$ extrapolations therefore describes
stabilization that cannot be inferred from convergence of a local charge or
dipole descriptor alone.

The magnitude of the local response is also strongly dependent on nucleobase
identity. AMP exhibits the largest monomer-centered dipole response, while
UMP shows the strongest response among the dimer-centered systems. GMP is
markedly weaker in both its fragment-charge redistribution and dipole
response, consistent with its comparatively weak size-dependent stabilization
in the energetic analysis. These differences show that the local electronic
response to chain growth is strongly modulated by the nucleobase-dependent
intermolecular environment.

\section{Conclusions}

In this work, we investigated the structural origin and size-dependent
stabilization of nucleotide 5$^\prime$-monophosphate crystals. Benchmarking
several density-functional approximations against the experimentally reported
triclinic, monoclinic, and orthorhombic CMP polymorphs identified PBE-D3(BJ)
as providing the most consistent description of the experimental lattice
parameters. The same method predicts the monoclinic polymorph to be the most
stable of the three, lying 1.76 and 3.00~kcal/mol per CMP below the
orthorhombic and triclinic forms, respectively. The energetic ordering cannot
be rationalized by H-bond count alone, pointing instead to the importance of
the topology and collective organization of the intermolecular packing.

A distinguishing feature of monoclinic CMP is an extended two-molecule
zig-zag motif propagating along the crystallographic $\vec{b}$ direction.
The ribose units participate in alternating intra- and intermolecular
O--H$\cdots$O interactions, while the nucleobase and phosphate regions form
part of the same repeating molecular arrangement. Extension to TMP, UMP,
AMP, and GMP shows that this chain-growth coordinate can be retained even
though the detailed H-bonding pattern, nucleobase--phosphate interactions,
protonation state, and lateral packing vary substantially across the series.

The COM calculations reveal a pronounced directional dependence of the
size-dependent stabilization. Successive dimer additions become increasingly
favorable along $\vec{b}$, whereas changes along $\vec{a}$ and $\vec{c}$ are
generally much smaller. This identifies propagation of the $\vec{b}$-direction
packing motif as the principal coordinate for cooperative growth. AMP shows
the strongest large-$N$ stabilization for the isolated COM assemblies,
whereas GMP is substantially weaker.

The COS calculations produce considerably greater stabilization than the
corresponding COM models. Because COS retains periodicity perpendicular to
the growth direction, this difference demonstrates the strong influence of
the lateral crystal environment on the propagating assembly. UMP exhibits
the strongest large-$N$ COS stabilization, while GMP again shows the weakest.
The change in ordering between COM and COS further shows that the lateral
environment does not provide a uniform additive contribution but modifies
the stabilization in a nucleobase-dependent manner.

The MBIS analysis provides a complementary view of the local electronic
response to chain growth. Fragment charges show substantial redistribution
between the nucleobase and phosphate regions for several NMPs, demonstrating
that the electronic response is not confined to the ribose H-bonding region.
Both the fragment charges and local dipole magnitudes respond strongly when
the environment beyond the directly interacting neighbors is introduced, but
subsequently approach saturation. This convergence describes the electronic
state of a fixed interior region and does not imply saturation of the
energetic stabilization of the growing assembly. Additional NMP units
continue to introduce intermolecular interactions, and finite COM and COS
assemblies remain energetically distinct from their corresponding large-$N$
limits even after the central electronic response has become nearly
converged.

Overall, our calculations indicate that stabilization along $\vec{b}$ arises
from the collective growth of an extended intermolecular packing motif in
which ribose H-bonding, base--phosphate organization, and the surrounding
crystal environment are coupled. The strong variation across the NMP series
shows that the magnitude of this size-dependent stabilization is controlled
not by a single H-bonding interaction but by the complete
nucleobase-dependent intermolecular environment.

\section*{CRediT authorship contribution statement}
\noindent 
{\bf PDD}: 
Conceptualization (equal); 
Analysis (equal); 
Data collection (equal); 
Writing (equal).
{\bf RR}: Conceptualization (equal); 
Analysis (equal); 
Data collection (equal); 
Funding acquisition; 
Project administration and supervision; 
Resources; 
Writing (equal).

\section*{Declaration of competing interest}
The authors declare that they have no known competing financial interests or personal relationships that could have appeared to influence the work reported in this paper.

\section*{Data Availability}
The data that support the findings of this study are within the article and its supplementary material.

\section*{Acknowledgements}
We acknowledge the support of the Department of Atomic Energy, Government of India, under Project Identification No.~RTI~4007. All calculations have been performed using the Helios computer cluster, which is an integral part of the MolDis Big Data facility, TIFR Hyderabad \href{http://moldis.tifrh.res.in}{(http://moldis.tifrh.res.in)}.

\appendix

\section*{References}
\bibliography{ref}






\end{document}



\newpage
\section*{Table of Contents}

\begin{enumerate}
    \item \textbf{Figure S1.} Mean absolute error (MAE) and standard deviation (SD) of the predicted lattice parameters for the triclinic, monoclinic, and orthorhombic CMP polymorphs obtained using the different DFT methods relative to experiment. Lattice constants and angles were considered in \AA{} and radians, respectively. (p. S3)
    \item \textbf{Table S1.} Computed structural parameters and relative energies of the triclinic, monoclinic, and orthorhombic CMP polymorphs. Lattice constants ($a$, $b$, and $c$) are given in \AA, angles ($\alpha$, $\beta$, and $\gamma$ ) in $^\circ$, volumes in \AA$^3$. Relative energies ($\Delta E$, in kcal/mol per formula unit) referenced to the corresponding monoclinic structure. (p. S4)
    \item \textbf{Table S2.} Hydrogen-bond geometry of the NMP crystals from experiment and PBE-D3(BJ) optimized structures, with distances $d$(D--H), $d$(H$\cdots$A) and $d$(D$\cdots$A) in \AA\ and the angle $\angle$(D--H$\cdots$A) in degrees. For the monoclinic ($P2_1$) structures (CMP, TMP, AMP) all molecules are symmetry-equivalent, so a single value is given for each interaction. For the triclinic ($P1$) structures of UMP and GMP, the two inequivalent molecules (A and B) of the AB dimer give two values per interaction; the value for molecule A is given first and that for molecule B in parentheses.(p. S5)
    \item \textbf{Table S3.} Electronic energies, $E_N$ (eV), of the chain-of-molecules
(COM) and chain-of-slabs (COS) models containing $N$ nucleotide
monophosphate units, extended along the crystallographic $\vec{a}$, $\vec{b}$, and $\vec{c}$ directions. (p. S6)
    \item Optimized crystal structures in POSCAR format: the triclinic, monoclinic, and orthorhombic structures of CMP, followed by those of TMP, UMP, AMP, and GMP. (p. S7--S20)
\end{enumerate}

\newpage
\begin{figure}[htbp]
    \centering
    \includegraphics[width=\columnwidth]{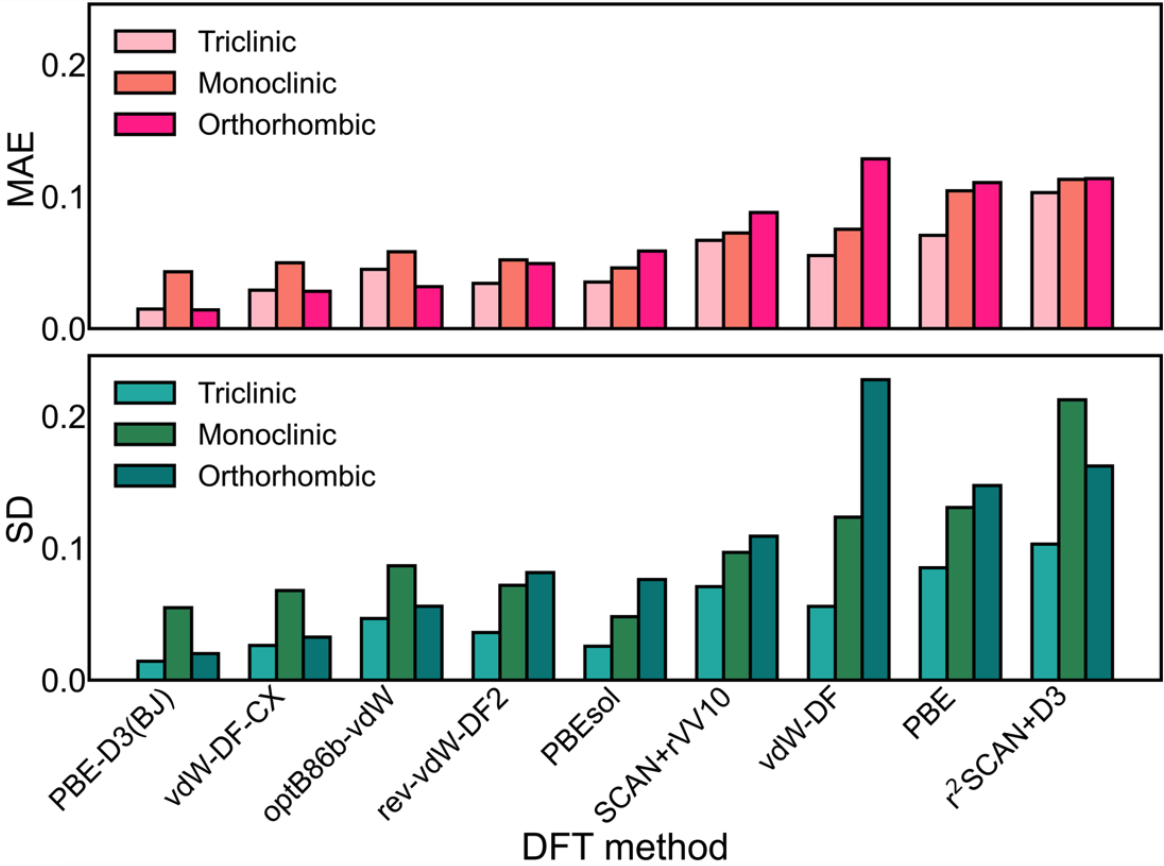}
    \caption{Mean absolute error (MAE) and standard deviation (SD) of the predicted lattice parameters for the triclinic, monoclinic, and orthorhombic CMP polymorphs obtained using the different DFT methods relative to experiment. 
    Lattice constants and angles were considered in \AA{} and radians, respectively. 
    }
    \label{fig:binding_energy}
\end{figure}

\begin{table*}[htbp]
\centering
\caption{Computed structural parameters and relative energies of the triclinic, monoclinic, and orthorhombic CMP polymorphs. Lattice constants ($a$, $b$, and $c$) are given in \AA, angles ($\alpha$, $\beta$, and $\gamma$ ) in $^\circ$, volumes in \AA$^3$. Relative energies ($\Delta E$, in kcal/mol per formula unit) referenced to the corresponding monoclinic structure.}
\vspace{2mm}

\renewcommand{\arraystretch}{1.15}
\begin{tabular}{llcccccccc}
\hline
Structure & Method &
$a$ & $b$ & $c$ &
$\alpha$ & $\beta$ &
$\gamma$ & $V$ & $\Delta E$\\
\hline

Triclinic $(Z=1)$ 
& $\mathrm{PBE\text{-}D3(BJ)}$
& 4.76 & 8.11 & 8.41
& 76.9 & 80.0 & 84.2
& 311.1 & 3.00 \\
& vdW\text{-}DF\text{-}CX & 4.76& 8.07 & 8.39 & 76.5 & 80.1 & 83.6 &307.5& 2.14\\
&$\mathrm{optB86b\text{-}vdW}$ & 4.74 & 8.00 & 8.39 & 76.0 & 79.7 & 83.1 &302.7& 1.16 \\
&$\mathrm{rev\text{-}vdW\text{-}DF2}$ &4.75&8.03&8.40&76.3&79.8&83.4&305.5&1.45 \\
&$\mathrm{PBEsol}$& 4.76 & 8.19 & 8.37 & 77.7 & 80.6 & 84.8 &313.8&4.93 \\
&$\mathrm{SCAN\text{+}rVV10}$& 4.70&7.95&8.34&76.2&79.8&83.3 &297.0 &0.72\\
&$\mathrm{vdW\text{-}DF}$&4.89&8.23&8.57&76.5&79.4&83.9 &328.9& 3.24\\
&$\mathrm{PBE}$& 4.88 & 8.36 & 8.50 & 77.7 & 80.3  & 85.1 &334.0&6.84 \\
&$\mathrm{r}^{2}\mathrm{SCAN\text{+}D3}$& 4.66 & 7.86 & 8.30& 75.6 & 79.2 & 82.4 &287.5&-1.53 \\
\hline

Monoclinic $(Z=2)$
& $\mathrm{PBE\text{-}D3(BJ)}$
& 4.82 & 8.57 & 15.86
& 90.0 & 92.7 & 90.0
& 654.1 & 0.00 \\
&vdW\text{-}DF\text{-}CX & 4.82 &8.54&15.89&90.0&93.3&90.0 &652.6&--\\
&$\mathrm{optB86b\text{-}vdW}$ & 4.81&8.43&15.83 &90.0&93.6&90.0  &640.5&--\\
&$\mathrm{rev\text{-}vdW\text{-}DF2}$ & 4.82&8.47 & 15.83 & 90.0 &93.4  &90.0  &644.8&--\\
&$\mathrm{PBEsol}$& 4.80 &8.77 &15.81 &90.0 & 92.1 &90.0  &664.6&--\\
&$\mathrm{SCAN\text{+}rVV10}$&4.77 &8.41&15.60&90.0&93.7&90.0  &624.6&--\\
&$\mathrm{vdW\text{-}DF}$& 4.94 &8.68&16.04&90.0&93.6&90.0  &686.9&--\\
&$\mathrm{PBE}$&4.92 &8.92&15.99&90.0 &92.8 &90.0  &700.2&--\\
&$\mathrm{r}^{2}\mathrm{SCAN\text{+}D3}$& 4.74 &8.11  &15.71&90.0& 94.8 &90.0 &601.6&--\\
\hline

Orthorhombic $(Z=4)$
& $\mathrm{PBE\text{-}D3(BJ)}$
& 8.33 & 31.36 & 4.85
& 90.0 & 90.0 & 90.0
& 1267.3 & 1.76 \\
&vdW\text{-}DF\text{-}CX & 8.28&31.39&4.85  &90.0&90.0&90.0 &1260.4&1.57\\
&$\mathrm{optB86b\text{-}vdW}$ &8.20 & 31.32 &4.84&90.0&90.0&90.0 &1242.8&0.60\\
&$\mathrm{rev\text{-}vdW\text{-}DF2}$ & 8.13 &31.36  &4.85  &90.0&90.0&90.0 &1248.8& 0.91\\
&$\mathrm{PBEsol}$& 8.47&31.15&4.85  &90.0&90.0&90.0 &1278.7&3.08\\
&$\mathrm{SCAN\text{+}rVV10}$& 8.15 & 31.07 &4.80&90.0&90.0&90.0&1215.4&0.29\\
&$\mathrm{vdW\text{-}DF}$& 8.43 &31.90&4.98 &90.0&90.0&90.0 & 1340.7&1.38\\
&$\mathrm{PBE}$& 8.64&31.62&4.96&90.0&90.0&90.0 &1354.2&3.78\\
&$\mathrm{r}^{2}\mathrm{SCAN\text{+}D3}$& 7.92&31.19  &4.75  &90.0&90.0&90.0&1173.6&0.55\\
\hline
\end{tabular}
\end{table*}

\clearpage
\begin{table*}[htbp]
\caption{Hydrogen-bond geometry of the NMP crystals from experiment and PBE-D3(BJ) optimized structures, with distances $d$(D--H), $d$(H$\cdots$A) and $d$(D$\cdots$A) in \AA\ and the angle $\angle$(D--H$\cdots$A) in degrees. For the monoclinic ($P2_1$) structures (CMP, TMP, AMP) all molecules are
symmetry-equivalent, so a single value is given for each interaction. For the triclinic ($P1$) structures
of UMP and GMP, the two inequivalent molecules (A and B) of the AB dimer give two values per interaction;
the value for molecule A is given first and that for molecule B in parentheses.}
\label{tab:hbond}
\renewcommand{\arraystretch}{1.2}
\footnotesize
\begin{tabular}{llllcccc}
\hline
NMP & Method & Type & D--H$\cdots$A & $d$(D--H) & $d$(H$\cdots$A) & $d$(D$\cdots$A) & $\angle$(D--H$\cdots$A) \\
\hline
CMP & Exp. & Intermolecular & O2$'$--H$\cdots$O3$'$ &0.98 &1.74 &2.67 &155 \\
    & PBE-D3(BJ) & Intermolecular & O2$'$--H$\cdots$O3$'$ &0.99 &1.64 &2.61 &163.8 \\
    &            & Intramolecular & O3$'$--H$\cdots$O2$'$ &0.99 &1.97 &2.60 &119.2 \\
\hline
TMP & PBE-D3(BJ) & Intermolecular  & O2$'$--H$\cdots$O3$'$ &0.99 &1.65 &2.62 &162.2 \\
    &            & Intramolecular & O3$'$--H$\cdots$O2$'$ &0.99 &2.00 &2.62 &118.1 \\
\hline
AMP & Exp. & Intermolecular  & O2$'$--H$\cdots$O3$'$ & 0.84&2.05 &2.81 &152 \\
    & PBE-D3(BJ) & Intermolecular  & O2$'$--H$\cdots$O3$'$ &0.99 &1.79 &2.74 &159.9 \\
\hline
UMP & PBE-D3(BJ) & Intermolecular  & O2$'$--H$\cdots$O3$'$ &0.99 (0.99)&1.64 (1.65) &2.60 (2.62) &162.5 (162.8) \\
    &            & Intramolecular & O3$'$--H$\cdots$O2$'$ &0.99 (0.99) &1.97 (1.96) &2.60 (2.60)&119.6 (120.0) \\
\hline
GMP & PBE-D3(BJ) & Intermolecular  & O2$'$--H$\cdots$O3$'$ &0.99 (0.99)&1.82 (1.73) &2.75 (2.68) &154.5 (160.0) \\
\hline
\end{tabular}
\end{table*}

\begin{table*}[ht]
\centering
\caption{Electronic energies, $E_N$ (eV), of the chain-of-molecules
(COM) and chain-of-slabs (COS) models containing $N$ nucleotide
monophosphate units, extended along the crystallographic $\vec{a}$, $\vec{b}$, and $\vec{c}$ directions.}
\label{tab:raw_energies}

\setlength{\tabcolsep}{5pt}

\begin{tabular}{cllrrrr}
\hline
Direction & NMP & Type &
\multicolumn{4}{c}{$N$} \\
\cline{4-7}
 & & & 2 & 4 & 6 & 8 \\
\hline

 \multirow{10}{*}{$\vec{a}$} & CMP & COM &$-$437.84330436 &$-$877.02335618 &$-$1316.19260258 & $-$1755.34268297 \\
          &     & COS &$-$443.19883978 & $-$890.79852465 &$-$1338.54458147 & $-$1786.30276217 \\
          & TMP & COM &$-$459.12549664 & $-$919.86929623& $-$1380.64595649 &$-$1841.40806414 \\
          &     & COS &$-$462.93612347 & $-$932.44709175 & $-$1402.03005100 &$-$1871.62718595 \\
          & UMP & COM & $-$425.40232980 &$-$852.32159433 & $-$1279.26188155 &$-$1706.18737703 \\
          &     & COS &$-$427.89223653 &$-$856.79686152 & $-$1285.76738562 & $-$1714.74244222 \\
          & AMP & COM &  $-$473.75950409&$-$948.93908851 & $-$1424.11009065 &$-$1899.26302534 \\
          &     & COS &$-$477.57249936 &$-$961.99797357 & $-$1446.73458840 &$-$1931.48994189 \\
          & GMP & COM &$-$490.54539351  &$-$981.74472390 & $-$1472.99095577 & $-$1964.24410798 \\
          &     & COS &$-$494.32488542 &$-$990.27906193 & $-$1486.45381416  & $-$1982.65352104 \\

\hline

\multirow{10}{*}{$\vec{b}$}  & CMP & COM & $-$437.84330436 &$-$878.59291451 &$-$1319.58183013 &$-$1760.63159112 \\
          &     & COS &$-$441.02043627 & $-$888.48685238 &$-$1336.87696709 & $-$1785.42181140  \\
          & TMP & COM &$-$459.12549664& $-$921.38446090 & $-$1383.86032765&$-$1846.39192560 \\
          &     & COS &$-$461.55120504 &$-$930.90050755 & $-$1401.00722333 & $-$1871.24464434 \\
          & UMP & COM &$-$425.40232998 & $-$853.92068163& $-$1282.64596800 &$-$1711.42489024 \\
          &     & COS &$-$425.38331114 &$-$859.18162380 & $-$1293.49319239 &$-$1727.89177551 \\
          & AMP & COM &$-$473.75950410 &$-$951.13599340 & $-$1428.79244142&$-$1906.51973322 \\
          &     & COS &$-$477.00229582& $-$961.52893014 &$-$1447.08869452 &$-$1932.81715733 \\
          & GMP & COM &$-$490.54539351 &$-$982.62286623 & $-$1474.79765161 & $-$1966.99798815\\
          &     & COS &$-$493.17626292 &$-$990.05400438 &$-$1487.35049983 &$-$1984.82778405 \\

\hline

\multirow{10}{*}{$\vec{c}$}  & CMP & COM &$-$437.84330378 &$-$876.72311172 &$-$1315.60220528 & $-$1754.48151578 \\
          &     & COS & $-$447.30475816 &$-$896.02023360 &$-$1344.73668350 &$-$1793.45286080 \\
          & TMP & COM &$-$459.12549664 &$-$918.30597521 & $-$1377.48417756 & $-$1836.66287361 \\
          &     & COS &$-$470.03548833 &$-$940.42220736 & $-$1410.81205797 &$-$1881.20226816 \\
          & UMP & COM &$-$425.40232980 &$-$850.66191559& $-$1275.91785608 &$-$1701.17461314 \\
          &     & COS &$-$434.42669797 &$-$868.23194946 & $-$1302.04170424 &$-$1735.85614756 \\
          & AMP & COM &$-$473.75950906 & $-$947.26164433 &$-$1420.75710240 & $-$1894.25280894 \\
          &     & COS &$-$484.53237788 &$-$970.33269089 & $-$1456.13316446 & $-$1941.93774169 \\
          & GMP & COM &$-$490.54539351 &$-$982.20569457 & $-$1473.87779019 &$-$1965.55131710 \\
          &     & COS &$-$494.28892508 & $-$992.74415083 &$-$1491.21150082 & $-$1989.68621865 \\

\hline
\end{tabular}
\end{table*}

\clearpage

\singlespacing
{
\setlength{\parskip}{0pt}
\setlength{\topsep}{0pt}
\setlength{\partopsep}{0pt}
\small

CMP-Triclinic \\
\vspace{-0.5\baselineskip}
\noindent{\ttfamily\small\leavevmode\xleaders\hbox{-}\hfill\kern0pt}

\begin{verbatim}
C9 H14 N3 O8 P
   1.00000000000000     
     4.7615052088327365   -0.0060160273097548   -0.0095412582935913
     0.8362686290530567    8.0702720796156875   -0.0026821216190784
     1.4759758048918954    1.7681183259515452    8.0916904591936465
   P    O    H    N    C
   1    8    14   3    9
Direct
  0.6801877665689826  0.8281208710427456  0.7029827112910095
  0.7656457977575168  0.0045497181995331  0.6912002436997290
  0.4011747800637823  0.7753825610491145  0.8123964950976141
  0.9299196789996911  0.6913478626450810  0.7605919222590390
  0.4073722183863833  0.1751088008230850  0.3680507497974052
  0.8675698968746194  0.4738674264775530  0.0593543079017029
  0.6116285386229717  0.8203923645747671  0.0784803753537684
  0.9656592644817321  0.5300929343877585  0.3807778853516010
  0.6465639971945155  0.7988182785311505  0.5244114011580607
  0.1208379758253071  0.7360055549239960  0.7718230349281233
  0.8533383868306306  0.5644901775591017  0.9589809409871918
  0.4852854763113095  0.7951846516904812  0.0059182686204693
  0.0505534193731795  0.0999742274142729  0.6325464842258667
  0.8573933942987724  0.0952156771350179  0.8927138613578873
  0.9436295208477663  0.2464321820303312  0.9964941511748040
  0.3771731355608460  0.4106738378637788  0.8889080049099596
  0.7266002107542747  0.5044305799839668  0.6321970873678613
  0.9225720566800302  0.2996477741376776  0.3265136630081079
  0.4530817278403315  0.4569583791860254  0.1923887750297448
  0.4468972464267134  0.6959069209341063  0.3165670536255327
  0.0758296341958351  0.7317555606796737  0.1814378968724126
  0.0847889274899949  0.8273343914484194  0.4405525267558320
  0.8462283783805800  0.9573918353432638  0.3144648475099392
  0.5728114887146083  0.3595180735215230  0.4979340971921867
  0.2002688906024351  0.1880585861462290  0.6324916471759212
  0.9933446216774071  0.1801738564745358  0.9041866272064593
  0.3956371821768075  0.2350101967597077  0.4913894874034302
  0.1853393520230937  0.2424611794630178  0.7752116184067459
  0.3784270248832415  0.3627188154508643  0.7784534672860237
  0.5672752461366817  0.4156046533410206  0.6388766749134671
  0.7909383677674572  0.4139349389369775  0.3490269333769574
  0.6599713091543519  0.5041532673009798  0.1952278474410242
  0.6310763517084957  0.6881872008030444  0.2182659270308940
  0.9026957281143763  0.7003366223774212  0.2864822928898961
  0.8840080070352918  0.8306410412500778  0.3914766912421815
\end{verbatim}
\noindent{\ttfamily\small\leavevmode\xleaders\hbox{-}\hfill\kern0pt}
}

\clearpage

\singlespacing
{
\setlength{\parskip}{0pt}
\setlength{\topsep}{0pt}
\setlength{\partopsep}{0pt}
\small

CMP-Monoclinic \\
\vspace{-0.5\baselineskip}
\noindent{\ttfamily\small\leavevmode\xleaders\hbox{-}\hfill\kern0pt}
\begin{verbatim}
C9 H14 N3 O8 P
   1.00000000000000
     4.8192035643470517    0.0000000000000000    0.0377380168880711
     0.0000000000000000    8.5650316636838326    0.0000000000000000
    -0.8819015486204730    0.0000000000000000   15.8385663132308121
   P    O    H    N    C
   2    16   28   6    18
Direct
  0.6416531286145309  0.7735951273432284  0.1502010710528498
  0.3583468713621704  0.2735951273663640  0.8497989289344039
  0.9047443847861771  0.7497395126592378  0.1035384024028374
  0.0952556152037972  0.2497395126823658  0.8964615976481382
  0.5366181219112179  0.9389293726176977  0.1579521125737233
  0.4633818780654765  0.4389293726408257  0.8420478874135301
  0.4105074115235194  0.6619964609069988  0.1080148191409408
  0.5894925884531760  0.1619964609301278  0.8919851808463120
  0.0130180771877835  0.0682632233539249  0.3344644517988665
  0.9869819227889105  0.5682632234464515  0.6655355481883930
  0.7328438740980381  0.3336284830645403  0.4897842921605592
  0.2671561258786636  0.8336284830414047  0.5102157078266938
  0.8913714702641021  0.6217850077304314  0.4730768942925753
  0.1086285297125911  0.1217850077535667  0.5269231056946774
  0.4390353282928300  0.4276547471460966  0.3193914174907175
  0.5609646716838644  0.9276547472386298  0.6806085824965353
  0.7075232152266705  0.7018216128443504  0.2433951581755372
  0.2924767847500312  0.2018216128674853  0.7566048418117226
  0.2092733917219153  0.6992551442173160  0.1112462381383455
  0.7907266082547790  0.1992551442404444  0.8887537618489142
  0.8586444383763354  0.2501513579213637  0.5132360246446904
  0.1413555616003592  0.7501513580138984  0.4867639753425700
  0.8557897140018978  0.5471809461948800  0.5189256520695595
  0.1442102859747957  0.0471809461023524  0.4810743479176931
  0.2660883488152841  0.0263336663914464  0.1986839311339668
  0.7339116511614096  0.5263336663683182  0.8013160688532934
  0.3603201666224249  0.0544565100988798  0.0695689082592483
  0.6396798333542690  0.5544565100757444  0.9304310917280051
  0.1719749273772758  0.1824027615087407  0.9963588432418943
  0.8280250725994184  0.6824027614856051  0.0036411567453664
  0.8048260955935428  0.3589886524160438  0.0547635881929310
  0.1951739043831512  0.8589886525085710  0.9452364117943217
  0.5634498749502006  0.4283523789271216  0.1860142669700896
  0.4365501250265021  0.9283523789039860  0.8139857330171701
  0.5201535339608186  0.2010530105667595  0.3581300715871112
  0.4798464660158749  0.7010530106592930  0.6418699284001490
  0.0428530064856625  0.3192883570910155  0.4032715440989878
  0.9571469934910319  0.8192883570678799  0.5967284558882643
  0.9666636384205293  0.5787814915704970  0.3489051049606989
  0.0333363615561649  0.0787814915936253  0.6510948950265534
  0.3908587681208316  0.5886125712764694  0.4164950505300155
  0.6091412318558695  0.0886125712996048  0.5835049494572446
  0.2855638314509325  0.7107158130872981  0.2718218381886493
  0.7144361685257613  0.2107158131104260  0.7281781617986032
  0.5181081874284187  0.8184136865348389  0.3409482490888290
  0.4818918125482757  0.3184136865579675  0.6590517508984242
  0.7789929390888178  0.2648819704593202  0.2606274633258346
  0.2210070609011634  0.7648819705518476  0.7393725367251391
  0.1123122084519702  0.1135415107700656  0.1968000115840536
  0.8876877915380039  0.6135415108625994  0.8031999884669196
  0.2071427307654720  0.1356337217101528  0.0564159450417220
  0.7928572692245024  0.6356337216870238  0.9435840550092524
  0.9700260108018440  0.1422975515053279  0.2682874576479611
  0.0299739891881312  0.6422975514822000  0.7317125424030204
  0.0600505623694525  0.1818489229706728  0.1198388221475612
  0.9399494376072418  0.6818489229475376  0.8801611778396919
  0.8532837558113434  0.2999016060856568  0.1141504709333794
  0.1467162441653518  0.7999016060625290  0.8858495290538808
  0.7214821284734487  0.3378656352345341  0.1853482736720602
  0.2785178715165326  0.8378656352114062  0.8146517263789221
  0.6305226977305842  0.3062683821423077  0.3383623135669486
  0.3694773022461105  0.8062683821191726  0.6616376864203108
  0.8329419942019192  0.3656817256315590  0.4089304266523113
  0.1670580057880614  0.8656817257240859  0.5910695733986715
  0.8222497277634349  0.5427040832443824  0.3964644948203417
  0.1777502722132594  0.0427040831518554  0.6035355051669191
  0.5241607958235693  0.5718364521980428  0.3628286256465123
  0.4758392041664043  0.0718364522211708  0.6371713744044613
  0.4934862499421384  0.7101725426118224  0.3038858809432683
  0.5065137500345553  0.2101725426349578  0.6961141190439916
\end{verbatim}
\noindent{\ttfamily\small\leavevmode\xleaders\hbox{-}\hfill\kern0pt}
}

\clearpage

\singlespacing
{
\setlength{\parskip}{0pt}
\setlength{\topsep}{0pt}
\setlength{\partopsep}{0pt}
\small

CMP-Orthorhombic \\
\vspace{-0.5\baselineskip}
\noindent{\ttfamily\small\leavevmode\xleaders\hbox{-}\hfill\kern0pt}
\begin{verbatim}
  C9 H14 N3 O8 P
   1.00000000000000     
     8.3326052391542351    0.0000000000000000    0.0000000000000000
     0.0000000000000000   31.3568939145114172    0.0000000000000000
     0.0000000000000000    0.0000000000000000    4.8504183318839624
   P    O    H    N    C
   4    32   56   12   36
Direct
  0.8376883917757516  0.1753376645644924  0.8944455325805565
  0.6623116081942635  0.8246623354195421  0.3944455327441167
  0.1623116081642785  0.6753376645724714  0.6055544673376672
  0.3376883917457668  0.3246623354115631  0.1055544672967749
  0.8138618380807735  0.1965999804114283  0.6183837850707639
  0.6861381620091741  0.8034000195726048  0.1183837850298728
  0.1861381619791963  0.6965999804194087  0.8816162150519042
  0.3138618380507881  0.3034000195965579  0.3816162150110119
  0.0076890960438104  0.1702253315921369  0.9977406538160338
  0.4923109039861738  0.8297746684238295  0.4977406537751433
  0.9923109040161594  0.6702253315681840  0.5022593461021898
  0.5076890959538557  0.3297746684158489  0.0022593462657491
  0.7296171408416542  0.2004377753559325  0.1116125814939264
  0.7703828591283609  0.7995622246281014  0.6116125815348182
  0.2703828592183093  0.7004377753639192  0.3883874185469655
  0.2296171408116688  0.2995622246201149  0.8883874185878499
  0.1223930616222688  0.0819604189519519  0.5748169072761543
  0.3776069384077088  0.9180395810640138  0.0748169074397140
  0.8776069384376939  0.5819604189599386  0.9251830926420693
  0.6223930615323212  0.4180395810560275  0.4251830926011779
  0.3712667212876111  0.0098238719883173  0.0650997659224558
  0.1287332786224415  0.9901761279957164  0.5650997659633473
  0.6287332786524272  0.5098238719963042  0.4349002341184288
  0.8712667213175950  0.4901761280196628  0.9349002341593208
  0.6828749412298516  0.0095748292941146  0.8397646228620752
  0.8171250587401635  0.9904251706899204  0.3397646228211846
  0.3171250587101783  0.5095748293021002  0.6602353770561484
  0.1828749413198062  0.4904251707138645  0.1602353770152559
  0.4761501213814239  0.0948811482111034  0.1796642569282412
  0.0238498786485542  0.9051188517729297  0.6796642569691337
  0.5238498785586068  0.5948811482190909  0.3203357431126501
  0.9761501214114083  0.4051188517649447  0.8203357431535344
  0.7563522514359646  0.1285569766985099  0.8696585379022083
  0.7436477485340505  0.8714430232855237  0.3696585380657678
  0.2436477486240049  0.6285569767064898  0.6303414620160083
  0.2563522514059804  0.3714430233094763  0.1303414619751239
  0.7633527962571777  0.1967908737064378  0.3124077625670246
  0.7366472037128374  0.8032091262775956  0.8124077626079163
  0.2366472038027913  0.6967908737144251  0.1875922372694156
  0.2633527962271943  0.3032091262696093  0.6875922373103073
  0.3998423554243775  0.9811947637749465  0.0015622749847436
  0.1001576446056074  0.0188052362410192  0.5015622751483030
  0.6001576446355922  0.4811947637669671  0.4984377249334728
  0.8998423553344300  0.5188052362170741  0.9984377248925815
  0.7274010404191218  0.0051218307372700  0.6514228779852909
  0.7725989596708258  0.9948781692467642  0.1514228779444076
  0.2725989596408471  0.5051218307452493  0.8485771219329257
  0.2274010403891380  0.4948781692707161  0.3485771218920328
  0.0912215610382360  0.1488414789308360  0.2842946854503959
  0.4087784389917486  0.8511585210851288  0.7842946854912805
  0.9087784389017943  0.6488414789388165  0.2157053145904959
  0.5912215610682208  0.3511585210771502  0.7157053144269361
  0.1258094271036667  0.2133368678724721  0.1551823289092540
  0.3741905728063856  0.7866631321434939  0.6551823289501449
  0.8741905728363721  0.7133368678485267  0.3448176711316307
  0.6258094271336430  0.2866631321355076  0.8448176711725232
  0.2556188851744159  0.2506802206806456  0.3298050839176561
  0.2443811147356287  0.7493197793033882  0.8298050839585481
  0.7443811147656135  0.7506802206886253  0.1701949161232357
  0.7556188852044016  0.2493197793273340  0.6701949161641200
  0.4319022535550719  0.2221073692439910  0.7163646325392227
  0.0680977464749129  0.7778926307400423  0.2163646324983391
  0.5680977465048982  0.7221073692519712  0.7836353673789939
  0.9319022534651240  0.2778926307639958  0.2836353673381007
  0.4895883473250155  0.1576351566394994  0.9993275580001799
  0.0104116527049694  0.8423648433764663  0.4993275579592886
  0.5104116527349544  0.6576351566155543  0.5006724421224882
  0.9895883472350679  0.3423648433684797  0.0006724420816034
  0.2411034576847030  0.0777302080786463  0.1057176450972869
  0.2588965423452824  0.9222697919373199  0.6057176451381718
  0.7588965423752666  0.5777302080547007  0.3942823549436046
  0.7411034577146810  0.4222697919293333  0.8942823549844963
  0.3557518728422849  0.0338629704510599  0.6616967288960105
  0.1442481271877001  0.9661370295329748  0.1616967288551271
  0.6442481270977455  0.5338629704590387  0.8383032712266576
  0.8557518728722696  0.4661370295249944  0.3383032711857647
  0.6216600952353363  0.0687008163498750  0.6741659190170021
  0.8783399047346859  0.9312991836660912  0.1741659189761106
  0.3783399048246347  0.5687008163259294  0.8258340809012216
  0.1216600952053502  0.4312991836581114  0.3258340810647807
  0.6425930831632000  0.0487410909498714  0.2839290620010506
  0.8574069168068151  0.9512589090341623  0.7839290618374974
  0.3574069168967690  0.5487410909578583  0.2160709380398343
  0.1425930831332155  0.4512589090261768  0.7160709380807261
  0.7708072442398464  0.1206890237241356  0.3008617106471496
  0.7291927557301758  0.8793109762598985  0.8008617106880408
  0.2291927557001927  0.6206890237321221  0.1991382893937354
  0.2708072443297919  0.3793109762519120  0.6991382894346273
  0.8771574986968748  0.0823251456259841  0.0997251727040140
  0.6228425012731403  0.9176748543899812  0.5997251725404548
  0.1228425012431554  0.5823251456339641  0.4002748273368777
  0.3771574986668970  0.4176748543820024  0.9002748273777689
  0.3259301337749517  0.1189461044709541  0.7953795667812440
  0.1740698662550332  0.8810538955450118  0.2953795667403503
  0.6740698662850099  0.6189461044470088  0.7046204333414241
  0.8259301338049377  0.3810538955370246  0.2046204333005343
  0.1796810728147237  0.1500418376653663  0.4376414682713433
  0.3203189272152612  0.8499581623186681  0.9376414681077833
  0.8203189272452457  0.6500418376733454  0.0623585317695489
  0.6796810727247694  0.3499581623426204  0.5623585318104333
  0.2078999755168155  0.2203458065061085  0.3072515712560063
  0.2921000245131696  0.7796541934779267  0.8072515710924456
  0.7921000245431475  0.7203458065140868  0.1927484287848858
  0.7078999755468001  0.2796541934699463  0.6927484288257780
  0.2031268513959377  0.1147630338053532  0.6019444594114498
  0.2968731486340468  0.8852369661786812  0.1019444593705653
  0.7968731486640251  0.6147630338133323  0.8980555405067667
  0.7031268514259225  0.3852369661707017  0.3980555406703261
  0.2516534070604120  0.1888742442079173  0.4711468395196432
  0.2483465929695734  0.8111257557761179  0.9711468395605347
  0.7483465928796186  0.6888742442158956  0.0288531605212419
  0.7516534070903965  0.3111257558000621  0.5288531605621334
  0.3697801961286699  0.1922418489188746  0.6798421361114857
  0.1302198039013080  0.8077581510651601  0.1798421362750462
  0.6302198039312918  0.6922418489268605  0.8201578638067308
  0.8697801960387233  0.3077581510891052  0.3201578637658382
  0.4018786479210103  0.1570760310006408  0.8342081817489138
  0.0981213521089749  0.8429239690153241  0.3342081819124734
  0.5981213521389600  0.6570760309766965  0.6657918181693099
  0.9018786478310622  0.3429239690073383  0.1657918181284190
  0.3537631155121426  0.0828399550793527  0.9928812989550672
  0.1462368843979096  0.9171600449046821  0.4928812989141758
  0.6462368844278881  0.5828399550873314  0.5071187009631565
  0.8537631155421270  0.4171600449286266  0.0071187011267160
  0.4100181078545836  0.0408908689595867  0.8626336191356049
  0.0899818920554620  0.9591091310563794  0.3626336190947200
  0.5899818920854468  0.5408908689675659  0.6373663809870632
  0.9100181078845683  0.4591091310483996  0.1373663809461712
  0.5920015504658597  0.0480278686894080  0.8496875225971194
  0.9079984495041554  0.9519721313265572  0.3496875225562343
  0.4079984494741687  0.5480278686654563  0.6503124775255487
  0.0920015504358760  0.4519721313185791  0.1503124774846572
  0.6239269635351614  0.0720406079925991  0.1179870894699461
  0.8760730365547862  0.9279593920233661  0.6179870895108381
  0.3760730365248017  0.5720406080005792  0.3820129103664940
  0.1239269635051763  0.4279593920153876  0.8820129104073855
  0.7667681781008940  0.1015774505381815  0.1104347459749966
  0.7332318218691211  0.8984225494777851  0.6104347460158885
  0.2332318219590691  0.6015774505461602  0.3895652540658954
  0.2667681780709166  0.3984225494698049  0.8895652541067867
\end{verbatim}
\noindent{\ttfamily\small\leavevmode\xleaders\hbox{-}\hfill\kern0pt}
}

\clearpage

\singlespacing
{
\setlength{\parskip}{0pt}
\setlength{\topsep}{0pt}
\setlength{\partopsep}{0pt}
\small

TMP-Optimized \\
\vspace{-0.5\baselineskip}
\noindent{\ttfamily\small\leavevmode\xleaders\hbox{-}\hfill\kern0pt}
\begin{verbatim}
  C10 H15 N2 O9 P
   1.00000000000000
     4.7569207866489425   -0.0011622794993635   -0.0625406355524772
     0.0011930460646267    8.6475240808940796   -0.0081976525663755
    -1.2502015345094684    0.0104711852370495   16.6046909445177562
   P    O    H    N    C
   2    18   30   4    20
Direct
  0.6292767868834803  0.7751062324983358  0.1632164635170160
  0.3744814987246485  0.2755451315097047  0.8367057767660834
  0.8972259312039980  0.7600857596848726  0.1235580576995241
  0.1068973283133913  0.2603510535816060  0.8765421063897495
  0.5132320766175376  0.9395009320817671  0.1709344829706759
  0.4905534268151456  0.4399745511820859  0.8291725735895216
  0.3932912546511090  0.6642588443889028  0.1233032544604153
  0.6107023552909913  0.1645102617907695  0.8762662633053139
  0.9779168184846515  0.0666558614172630  0.3514366964833603
  0.0246297914478710  0.5670525236158404  0.6485725932010946
  0.7278860124494244  0.3480870316741479  0.4905425896672809
  0.2739045795158266  0.8481529781156865  0.5093840906425098
  0.8990733653722301  0.6321637416863295  0.4712805244294258
  0.1026699664407843  0.1322471926778303  0.5286369333537235
  0.4289681616050160  0.4352990591211721  0.3230158310212933
  0.5734861322635385  0.9358960973162591  0.6768357353057948
  0.6966734952594834  0.7114889057908741  0.2545340728345555
  0.3061508102289141  0.2122706071839536  0.7453276097118154
  0.2140424783999663  0.1244843531261840  0.0922972562103686
  0.7902808798568789  0.6248296615655394  0.9078931377395312
  0.1856971943244756  0.7016665754194772  0.1248541751597295
  0.8183458330002317  0.2020527787185449  0.8748627465921476
  0.8517408126056627  0.2644333980747706  0.5148768021830364
  0.1499463927000130  0.7644666666327199  0.4850850811164939
  0.8695345465480209  0.5617885870619908  0.5165849482932058
  0.1320907062705947  0.0617851819636787  0.4833399684720414
  0.3551297013878446  0.0342859821858130  0.1170085556635173
  0.6491014816538881  0.5347516760601079  0.8832084521494658
  0.4257349949820244  0.9331852974039748  0.8014532916803502
  0.5049748949234110  0.2125014525063424  0.3653370955221738
  0.4977554569035786  0.7130078729136188  0.6345451806789180
  0.0399482513481734  0.3275066029780997  0.4112798783501929
  0.9623533652558305  0.8276826377076658  0.5888166270564353
  0.9608733258898373  0.5811668721451542  0.3536695704318588
  0.0414168512632409  0.0815103007346210  0.6462982357558025
  0.3825026128325604  0.5937558892214607  0.4159225061445237
  0.6194059324315394  0.0940681058364526  0.5838156750668636
  0.2673714417412457  0.7129722771448754  0.2761546912122094
  0.7352319617880917  0.2138207547475762  0.7234537267610633
  0.4988171505982802  0.8220943861733864  0.3454400914718804
  0.5031820238836786  0.3226142080267692  0.6542255969654509
  0.2865359954853166  0.7899115493573259  0.9888964552822606
  0.9850112296463884  0.9064268838771646  0.9677194404543888
  0.3194734814721057  0.9741214916866365  0.9418207646396053
  0.5777443616588225  0.4327118901468379  0.1985175823837312
  0.7637958677367147  0.5135174750383253  0.7729935804842275
  0.2394198915861626  0.0130386565087747  0.2271438908435253
  0.6836101196248601  0.4731046589381758  0.0581040824573793
  0.7206449229063711  0.2892333142723221  0.0109452108812365
  0.0199333513524397  0.4071691279798043  0.0328299439049771
  0.7687413241254479  0.2652885219624995  0.2743568890877200
  0.2342777500745063  0.7657339648292146  0.7256106249932432
  0.0958314925489481  0.1037424300074243  0.2220982282776994
  0.9075891198930224  0.6041663307264599  0.7780016508956830
  0.8194622915785322  0.3715708496663210  0.0551179228758744
  0.1853476319554341  0.8719182389482429  0.9449753472379878
  0.9477686706863628  0.1388676408913698  0.2872194155763564
  0.0551980079787528  0.6392875678301996  0.7128048792511390
  0.0629762142158375  0.1726839974060197  0.1482696408268648
  0.9410023267051755  0.6730644878991995  0.8518501514137120
  0.8674638291582187  0.2962071682390104  0.1358379817307397
  0.1365915194463297  0.7965839613512216  0.8642302145167429
  0.7293083352305891  0.3381202623196807  0.2014222424458119
  0.2742156178026836  0.8385468196886103  0.7985636511311806
  0.6188843761172317  0.3143692836806155  0.3455812162107363
  0.3837179659028543  0.8148172156692952  0.6543303039100961
  0.8276823408592501  0.3751529193035498  0.4141055613649984
  0.1745339883950285  0.8753675405994155  0.5858683089396773
  0.8193906131955576  0.5496038854439057  0.3994458553431343
  0.1827052304113870  0.0498710128221894  0.6004792679687145
  0.5149006330730380  0.5782196980830004  0.3655559654542873
  0.4872836657392154  0.0786697545730963  0.6342505882015973
  0.4784434813718273  0.7153262965788250  0.3092244257729092
  0.5239462672498261  0.2159762767370851  0.6904997697519203
\end{verbatim}
\noindent{\ttfamily\small\leavevmode\xleaders\hbox{-}\hfill\kern0pt}
}

\clearpage

\singlespacing
{
\setlength{\parskip}{0pt}
\setlength{\topsep}{0pt}
\setlength{\partopsep}{0pt}
\small

UMP-Optimized \\
\vspace{-0.5\baselineskip}
\noindent{\ttfamily\small\leavevmode\xleaders\hbox{-}\hfill\kern0pt}
\begin{verbatim}
 C9 H13 N2 O9 P
   1.00000000000000
     4.7761235521636483    0.0121858129721556    0.2371443152171647
     0.0260138356474203    8.4062837480286365   -0.9276604285631458
    -0.2102549049752460    0.0949203696295333   15.4931105322209621
   P    O    H    N    C
   2    18   26   4    18
Direct
  0.7541293015637608  0.6916328687263436  0.1318186791709758
  0.2175621456420876  0.3495314434265040  0.8674186017696627
  0.0333566160884217  0.6633343715377217  0.0865156106709211
  0.9371785190436432  0.3362146282620785  0.9117764812016175
  0.6357723604359357  0.8615206670848798  0.1372222510554330
  0.3268767586169746  0.5175460540176809  0.8579952397393352
  0.5289458366578025  0.5663555291293170  0.0979419992870655
  0.4396794115363376  0.2464563875401529  0.9087321462408039
  0.0420331413324933  0.0295306819694848  0.3373291215131898
  0.9541175124209497  0.5937234100311433  0.6668868852554491
  0.7270938883125939  0.3415639892790846  0.4886971288752989
  0.2695769564711064  0.8464033155487256  0.5137413572218390
  0.9009296362651503  0.6268150982989367  0.4689858417707495
  0.0959053056193677  0.1398612226147206  0.5316800593396408
  0.4948540979406862  0.3964954769512503  0.3100641017654673
  0.4945627117797284  0.9732814034636588  0.6931640581709709
  0.7955694254123890  0.6576740760581665  0.2321097372681724
  0.1878785325056880  0.2657179644725821  0.7688061614713751
  0.3539818027888448  0.0275257322042666  0.0545696349724046
  0.5756624981314615  0.7326142479804273  0.9400106724649293
  0.3249530539346103  0.6064840980790445  0.0947535388374575
  0.6451160861372596  0.2856091296333308  0.9085598604721594
  0.8453608826897925  0.2637961984463152  0.5153357524336762
  0.1520654366263005  0.7585648817018190  0.4867263905553669
  0.8476364392765495  0.5613215453224779  0.5163021762364660
  0.1483120369185695  0.0559458265178731  0.4845191100562375
  0.4888876418856656  0.9401176148151058  0.0803834721377729
  0.6252928787501945  0.5760066019267551  0.7994131102438921
  0.9818104881638587  0.2634551448097354  0.0436030946269702
  0.9620169712203492  0.9625204821208997  0.9528600775185311
  0.6819910819596611  0.3659870553005228  0.1736436174124917
  0.2875992408072011  0.0045207404330510  0.8271668861268451
  0.5561875119007667  0.1771916481978801  0.3546169649744980
  0.4333392222053060  0.7360203765510289  0.6506331610326188
  0.0697106176795437  0.3043494922100159  0.4023874555521759
  0.9241585507111915  0.8435304569460346  0.5988730251218426
  0.0165509790503799  0.5538384283247315  0.3420090947089539
  0.9734925816174529  0.1163301370806674  0.6581661704555577
  0.4143103950746089  0.5809722416451852  0.4083810948063145
  0.5781928697991345  0.1199536905242969  0.5948415776662850
  0.3603416045287945  0.6702720168949609  0.2574832634722491
  0.6242918942943897  0.2682168470601564  0.7456050064164091
  0.5690213714339574  0.7957082003929782  0.3297653174315836
  0.4200391352092454  0.3646308507357238  0.6715312255357087
  0.4492753539136776  0.6322957581318186  0.9159286120400513
  0.3440534363222985  0.9481414295565569  0.2009749453393674
  0.8525462398068973  0.2138772202913175  0.2552240138901608
  0.1298443515676788  0.8153485403599688  0.7470730804030752
  0.2008064154125445  0.0390185857614974  0.1964829620587379
  0.7689587833245620  0.6686577442892195  0.8037090516415584
  0.0314951955115035  0.0887040347298406  0.2679255733008533
  0.9525482131523294  0.6844379013696811  0.7346134193880220
  0.1896087725125468  0.0921656247632260  0.1162820899662362
  0.7612478715799155  0.7617769328833588  0.8811718555416093
  0.9980495270693379  0.2151179870016367  0.1057842824381472
  0.9543998914003040  0.8873176947335971  0.8920407134916021
  0.8353637119247416  0.2717724940688565  0.1766676561515091
  0.1330289899402023  0.9099326411191954  0.8236870072977135
  0.6763609548081583  0.2776874855137365  0.3330566051755386
  0.3133510112741673  0.8455390343409119  0.6702083886080763
  0.8574184873948127  0.3536788440703635  0.4060014947200409
  0.1365701248761025  0.8915691594040045  0.5957520445986582
  0.8551003811897355  0.5294156702719870  0.3907037000277374
  0.1377010545525493  0.0733122185831230  0.6103878297138572
  0.5668534139003679  0.5519748109619891  0.3541906360114868
  0.4233453358751426  0.1114237515133031  0.6483876506905131
  0.5581247137416195  0.6781436048483001  0.2920211609425988
  0.4285523380598701  0.2622680146715777  0.7099790137836364
\end{verbatim}
\noindent{\ttfamily\small\leavevmode\xleaders\hbox{-}\hfill\kern0pt}
}

\clearpage

\singlespacing
{
\setlength{\parskip}{0pt}
\setlength{\topsep}{0pt}
\setlength{\partopsep}{0pt}
\small

AMP-Optimized \\
\vspace{-0.5\baselineskip}
\noindent{\ttfamily\small\leavevmode\xleaders\hbox{-}\hfill\kern0pt}
\begin{verbatim}
C10 H14 N5 O7 P
   1.00000000000000
     4.8080181652886820    0.0000000000000000   -0.0121817979495526
     0.0000000000000000   10.2492482570965251    0.0000000000000000
    -0.2771495265043318    0.0000000000000000   13.6613803028287428
   P    O    H    N    C
   2    14   28   10   20
Direct
  0.4316268369501969  0.2557277505852434  0.1522483878973752
  0.5683731630708989  0.7557277506338798  0.8477516121099217
  0.1454180903745022  0.2772858231222915  0.1088025397744793
  0.8545819096466009  0.7772858230736559  0.8911974602328188
  0.5336919791847177  0.1163763173108533  0.1592046211446318
  0.4663080208363770  0.6163763173594904  0.8407953788626644
  0.6411480268378205  0.3468833803767313  0.0963447800858695
  0.3588519731832760  0.8468833804253679  0.9036552199214278
  0.6954675258454404  0.6565343137746192  0.5169840377544007
  0.3045324741756559  0.1565343138232627  0.4830159622528961
  0.3591315654760808  0.4211233908725251  0.5075620570641853
  0.6408684345450216  0.9211233908238884  0.4924379429431116
  0.8153681503928771  0.5386300977492928  0.3165610504640292
  0.1846318496282187  0.0386300977006564  0.6834389495432670
  0.4256047588910494  0.3157988578322194  0.2627912721717980
  0.5743952411300532  0.8157988578808559  0.7372087278354915
  0.8438536287622765  0.3175798004079400  0.1024801224062439
  0.1561463712588188  0.8175798003593034  0.8975198776010462
  0.7064010156575387  0.7533489030615609  0.5207219391725917
  0.2935989843635651  0.2533489031102045  0.4792780608347058
  0.4990059335169473  0.4237926603559528  0.5636481798295655
  0.5009940665041553  0.9237926604045890  0.4363518201777302
  0.7948826677101123  0.0574461940099063  0.2080737916481961
  0.2051173323109841  0.5574461939612695  0.7919262083590938
  0.6875320639153032  0.9890031174795076  0.0644993735136534
  0.3124679358948110  0.4890031175281437  0.9355006264206924
  0.8372463958867020  0.8502622168695911  0.0073268235984810
  0.1627536041343944  0.3502622168209470  0.9926731764088081
  0.0756708390884415  0.0630595326834244  0.3570165297196380
  0.9243291609326546  0.5630595326347873  0.6429834702876599
  0.5556444002765146  0.5418628693924231  0.1511477393697165
  0.4443555997445810  0.0418628694410603  0.8488522606375809
  0.8347310836952169  0.7344613974432717  0.3506982439347200
  0.1652689163258791  0.2344613973946355  0.6493017560725770
  0.3276025525044877  0.6918284151821066  0.4372483940877144
  0.6723974475129207  0.1918284151334697  0.5627516058466382
  0.2855622203587128  0.4795615014366383  0.3677805962289379
  0.7144377796623838  0.9795615014852754  0.6322194037783596
  0.8784317189041175  0.4161599210411603  0.4372631967004937
  0.1215682811169853  0.9161599210897967  0.5627368033068034
  0.8625735277510501  0.2978754108864292  0.2730876785467550
  0.1374264722700457  0.7978754108377922  0.7269123214605424
  0.6699291642154590  0.2258856016190378  0.3700459694428599
  0.3300708358056439  0.7258856016676745  0.6299540305644304
  0.9553049909671621  0.9861284418800618  0.2214776665794445
  0.0446950090502389  0.4861284419286980  0.7785223333549011
  0.3018252533591728  0.8963793822365203  0.3267452723090540
  0.6981747466619230  0.3963793821878849  0.6732547276982436
  0.8222729128563822  0.9114908660077173  0.0672971797559791
  0.1777270871647142  0.4114908659590804  0.9327028202513175
  0.2581482844348795  0.6906140303327762  0.1127890254302115
  0.7418517155862170  0.1906140303814202  0.8872109745770854
  0.4940922423125899  0.6954886235123874  0.2573709677856675
  0.5059077577085137  0.1954886235610243  0.7426290322216291
  0.1114087695026655  0.9852292294340393  0.3045961881825850
  0.8885912305147424  0.4852292294826826  0.6954038117517600
  0.3261711120151918  0.8039520139103230  0.2558751955591186
  0.6738288880059113  0.3039520139589597  0.7441248044481712
  0.1806212722413359  0.7984509671678516  0.1656167284893328
  0.8193787275724734  0.2984509672164883  0.8343832715179644
  0.9820425579527983  0.8978359445667784  0.1461051128239892
  0.0179574420682981  0.3978359445181424  0.8538948871833014
  0.4459241448452771  0.6311392481444728  0.1690907090437037
  0.5540758551758191  0.1311392481931098  0.8309092909635872
  0.6791649306904403  0.6567296119001015  0.3395799520191485
  0.3208350693306549  0.1567296118514650  0.6604200479881485
  0.5184045131938865  0.6327354336459855  0.4344067266705964
  0.4815954868272111  0.1327354336946225  0.5655932733367007
  0.4529197798768348  0.4875809109341733  0.4235511249930136
  0.5470802201442606  0.9875809109828106  0.5764488750142838
  0.7191181764825526  0.4335599493589259  0.3792101112687271
  0.2808818235385429  0.9335599494075633  0.6207898887385698
  0.6808391984191851  0.3098208732704007  0.3199319070527580
  0.3191608016019121  0.8098208733190448  0.6800680929545322
\end{verbatim}
\noindent{\ttfamily\small\leavevmode\xleaders\hbox{-}\hfill\kern0pt}
}

\clearpage

\singlespacing
{
\setlength{\parskip}{0pt}
\setlength{\topsep}{0pt}
\setlength{\partopsep}{0pt}
\small

GMP-Optimized \\
\vspace{-0.5\baselineskip}
\noindent{\ttfamily\small\leavevmode\xleaders\hbox{-}\hfill\kern0pt}
\begin{verbatim}
 C10 H14 N5 O8 P
   1.00000000000000     
     5.1516913323585118   -0.2011644163354339   -0.3671186795699342
    -0.0674982460221634   10.3181260097561633    0.1869606558353626
    -1.2613152829999672   -0.1517754737607795   13.2731756322386030
   P    O    H    N    C
   2    16   28   10   20
Direct
  0.5315600002525840  0.2444409985402511  0.1555038229631709
  0.5279531426235075  0.7498537168560951  0.8345308515589996
  0.2292668228084314  0.2340611694057514  0.1160125660388826
  0.7847673864811754  0.7070200101152111  0.8168227585090554
  0.6646970549992312  0.1208136364136299  0.1482567875727237
  0.3756324590797876  0.6502755332822313  0.8974806038094875
  0.6532786096205468  0.3700511886751423  0.1130647174920163
  0.5291209558629711  0.8863542257181218  0.8855712088339076
  0.6416992973611418  0.6417198456793796  0.5319031771305122
  0.3329924332331248  0.1297696185693740  0.4654551232034406
  0.3969963630156056  0.3943645556588062  0.4878141428166256
  0.5249228512847204  0.8925520015976673  0.5267159742392343
  0.8703101114493962  0.5443686459885961  0.3441804731335551
  0.0599826075472894  0.0660729015669786  0.6449999700280550
  0.5311053383263538  0.2826198773700670  0.2726733353807447
  0.3412757228532865  0.7594506098885083  0.7235979066452288
  0.1635237087328974  0.9275547126439975  0.9879484481020594
  0.9781056396802492  0.4013554615826922  0.9945790631247913
  0.7795462929542801  0.3688444796597894  0.0602077875297615
  0.3812554211078987  0.8995125371539954  0.9283663023722278
  0.6014086003284282  0.7323130974826813  0.5465355432453204
  0.3443999726245107  0.2244543896455563  0.4518409715768798
  0.5142126545965058  0.4035038200702671  0.5556320459779001
  0.4348421364438583  0.9086936717722928  0.4565480514564512
  0.9334094364275490  0.0214995655508920  0.1218626153874197
  0.4370444762915638  0.6673108417299736  0.9777076453112877
  0.7738556644232853  0.5702753674007932  0.1614060504057195
  0.1722790283238106  0.1011772211692070  0.8394249404421515
  0.8598123603413177  0.7353800101894945  0.3901743988397232
  0.1376364370553733  0.2516896836742726  0.5978727285973072
  0.3418622842995468  0.6632134801022996  0.4087931339937096
  0.6328426610898560  0.1656649341024072  0.5884908333640937
  0.3636927813158303  0.4516412930972306  0.3410145374156973
  0.5804250452175673  0.9631288370948368  0.6701152453886104
  0.9054670966148958  0.4246365056871980  0.4709127425204381
  0.0359476511017593  0.9101263538784229  0.5467866646022626
  0.9436418242950489  0.3099374941003550  0.2977705821313887
  0.1128628244629413  0.9005699994463262  0.7731663573342378
  0.8003630019810376  0.2279093701617947  0.3951707991653172
  0.9454421206701544  0.7862625504083351  0.6862686173280691
  0.1745349712410713  0.5899078821637630  0.6977481889724985
  0.9200436417988046  0.5555523814318297  0.6019635542927563
  0.0578213165575390  0.0316866174927747  0.3775577928994866
  0.8364194228673019  0.0598803585200070  0.2650970839792569
  0.1425953515884415  0.5223801687417085  0.8563005478276795
  0.1581037887161686  0.2991608636389267  0.0634097425016259
  0.9673765730537752  0.0004400289400576  0.3074027601559851
  0.0383551415880833  0.5210649388214939  0.6650584309841813
  0.0644920955542230  0.9514418940179846  0.1486406607293792
  0.9991623865615781  0.4512701009835076  0.8287098000843808
  0.3045275038657622  0.8608374685573773  0.2998336546187744
  0.6976427820453589  0.3733881638981387  0.6848346241389152
  0.5097931140713652  0.7082894644811549  0.0867604985910107
  0.4954700050694307  0.2251906056731051  0.9066352247753870
  0.5961425972671069  0.6962254516227137  0.2561822563604110
  0.3786559509140812  0.2132720564913188  0.7360462688115276
  0.1186804067930297  0.9365569984980262  0.2526190655148540
  0.9060029330125529  0.4486278655354424  0.7262084492901862
  0.4158362200154515  0.7917944318624054  0.2337564620158284
  0.5863289046452929  0.3033392896370652  0.7538281286815322
  0.3660577745288627  0.7980007736007350  0.1275090631479463
  0.6590371832886491  0.3068769892400106  0.8600473100310616
  0.1969170081289846  0.8933068111001092  0.0792996587216351
  0.8815446684612931  0.3855465425250564  0.9008197873559846
  0.6464735332545669  0.6495611016921546  0.1650256373571904
  0.3308914689284682  0.1710134472469867  0.8313954066869912
  0.7246866576615347  0.6551928032264810  0.3572796865642774
  0.2432816233507899  0.1694984864811261  0.6350559891126926
  0.5288156263083978  0.6158584987112526  0.4294548204552391
  0.4367205630589719  0.1179413633204933  0.5705838416992458
  0.4986820421041411  0.4692535547505200  0.4126683883018027
  0.4404545492855418  0.9739389466785401  0.5994555005410820
  0.7736840921471082  0.4338389368060506  0.3975034144354269
  0.1591775293221392  0.9426372299595476  0.6183646209387604
  0.7767782048757069  0.3076299301010906  0.3396398420602732
  0.1266720931371530  0.8463586719290437  0.7041243121292362
\end{verbatim}
\noindent{\ttfamily\small\leavevmode\xleaders\hbox{-}\hfill\kern0pt}
}